\documentclass[floats,floatfix,amssymb,prd,twocolumn,superscriptaddress,nofootinbib,preprintnumbers]{revtex4-1}
\pdfoutput=1
\usepackage{subcaption,ragged2e}
\DeclareCaptionJustification{justified}{\justifying}
\usepackage{amssymb,amsmath,verbatim,mathtools,needspace,enumitem,etoolbox,graphicx,microtype,afterpage,bm}
\usepackage{tikz}
\usetikzlibrary{shapes.misc, positioning, arrows.meta}
\usepackage{framed}

\usepackage{hyperref}
\definecolor{linkcolor}{rgb}{0.0, 0.47, 0.75}
\definecolor{citecolor}{rgb}{1.0, 0.5, 0.0}
\hypersetup{
  linkcolor  = linkcolor,
  citecolor  = linkcolor,
  urlcolor   = linkcolor,
  colorlinks = true
}
\usepackage{multirow,pifont,lmodern,float,tabularx,booktabs}
\usepackage[all]{hypcap}
\usepackage[T1]{fontenc}
\usepackage[utf8]{inputenc}
\usepackage{appendix}
\usepackage{cleveref}

\allowdisplaybreaks

\newcommand{\datavec}{x}
\newcommand{\thetavec}{\theta}
\newcommand{\xobs}{x_{\rm{obs}}}

\DeclarePairedDelimiterX{\PQ}[2]{(}{)}{%
  #1\,\delimsize\|\,#2%
}
\newcommand{\KL}{{\rm KL} \, \PQ}

\begin{document}

\title{Simulation-based inference with deep ensembles:\\ Evaluating calibration uncertainty and detecting model misspecification}

\author{James Alvey}
\email{jbga2@cam.ac.uk}
\thanks{ORCID: \href{https://orcid.org/0000-0003-2020-0803}{0000-0003-2020-0803}}
\affiliation{Kavli Institute for Cosmology Cambridge, Madingley Road, Cambridge CB3 0HA, United Kingdom}
\affiliation{Institute of Astronomy, University of Cambridge, Madingley Road, Cambridge CB3 0HA, United Kingdom}

\author{Carlo R. Contaldi}
\email{c.contaldi@imperial.ac.uk}
\thanks{ORCID: \href{https://orcid.org/0000-0001-7285-0707}{0000-0001-7285-0707}}
\affiliation{Department of Physics, Imperial College London, SW7 2AZ, London, United Kingdom}

\author{Mauro Pieroni}
\email{mauro.pieroni@csic.es}
\thanks{ORCID: \href{https://orcid.org/0000-0003-0665-266X}{0000-0003-0665-266X}}
\affiliation{Theoretical Physics Department, CERN, 1211 Geneva 23, Switzerland}
\affiliation{Instituto de Estructura de la Materia (IEM), CSIC, Serrano 121, 28006 Madrid, Spain}

\begin{abstract}
\noindent Simulation-Based Inference (SBI) offers a principled and flexible framework for conducting Bayesian inference in any situation where forward simulations are feasible. 
However, validating the accuracy and reliability of the inferred posteriors remains a persistent challenge. 
In this work, we point out a simple diagnostic approach rooted in ensemble learning methods to assess the internal consistency of SBI outputs that does not require access to the true posterior. 
By training multiple neural estimators under identical conditions and evaluating their pairwise Kullback-Leibler (KL) divergences, we define a consistency criterion that quantifies agreement across the ensemble. 
We highlight two core use cases for this framework: a) for generating a robust estimate of the systematic uncertainty in parameter reconstruction associated with the training procedure, and b) for detecting possible model misspecification when using trained estimators on real data. 
We also demonstrate the relationship between significant KL divergences and issues such as insufficient convergence due to, e.g., too low a simulation budget, or intrinsic variance in the training process. 
Overall, this ensemble-based diagnostic framework provides a lightweight, scalable, and model-agnostic tool for enhancing the trustworthiness of SBI in scientific applications.
\end{abstract}

\maketitle

\section{Introduction}

\noindent Simulation-Based Inference (SBI) has emerged as a powerful framework to perform Bayesian inference in diverse scientific domains (see, e.g.,~\cite{cranmer2020frontier} for a review). 
In the broadest terms, the range of SBI algorithms, such as Neural Posterior Estimation (NPE)~\cite{greenberg2019automatic, papamakarios2016fast, lueckmann2021benchmarking}, Neural Ratio Estimation (NRE)~\cite{hermans2020likelihood,rozet2021arbitrary}, or Neural Likelihood Estimation (NLE)~\cite{Alsing:2019xrx} (as well as their sequential analogues, see, e.g.,~\cite{Papamakarios2019sbi,Miller:2021hys,deistler2022truncated}), all leverage machine learning (ML) techniques to approximate various statistical quantities relevant for Bayesian inference. 
All these approaches are based on the basic ability to sample parameters $\theta$ from some prior $p(\theta)$ and subsequently generate simulated data $x \sim p(x | \theta)$.

Despite the increasing sophistication and adoption of SBI methods, a fundamental challenge persists: how do we assess the quality and reliability of the inferred posterior distribution $q(\theta|\xobs) \approx p(\theta|\xobs)$ for a specific observed dataset $\xobs$? Since the true posterior is (at least in the most well-motivated use cases) often unknown and/or analytically intractable, direct comparison is rarely possible. 
Without robust validation methods, it remains challenging to determine whether an SBI result is accurate, whether the algorithm has converged sufficiently, or whether the resulting posterior approximation accurately captures the true uncertainty about the parameters. 
This lack of ground-truth comparison hinders the confident application and interpretation of SBI results in scientific discovery.

Consequently, there is a clear and pressing need for a suite of reliable diagnostics and validation techniques specifically tailored for SBI outputs. 
Several valuable approaches have already been developed to address this need. 
Simulation-Based Calibration (SBC)~\cite{cook2006validation, Talts:2018zdk, hermans2021trust} provides a necessary condition for correctness by checking whether, on average, the ranks of the true parameters under the inferred posteriors are uniformly distributed over many simulated datasets. 
Probability-Probability (PP) plots are often used to assess this calibration. 
More recently, techniques like Tests of Accuracy with Random Points (TARP)~\cite{lemos2023sampling} have been introduced to quantify the concentration and calibration of posteriors by measuring the coverage probability of credible regions. 
Standard Bayesian techniques, such as posterior predictive checks, can also be adapted. 
Broadly, this involves comparing simulations generated from the synthetic model to the observed data to check for model misspecification or goodness-of-fit. 
A variety of methods have been developed to address this challenge, including approaches that learn statistics to distinguish between data from different models~\cite{AnauMontel:2024flo,cranmer2016approximating, dalmasso2020confidence, heinrich_learning_2022, huang_learning_2023, wehenkel_addressing_2024}, adaptations of Bayesian model comparison and evidence estimation~\cite{jeffrey_evidence_2024, Gessey-Jones:2023vuy, mancini_bayesian_2023}, deviations in different SBI algorithms~\cite{Cannon:2022aaa,schmitt_detecting_2024-1,Kelly:2025aaa}, and diagnostics tailored to specific scientific applications~\cite{von_wietersheim-kramsta_kids-sbi_2024, Pierre:2025ulp}.
In general, while these methods are extremely valuable, they often assess average performance over many simulations (like SBC), or rely on visual inspection (like PP-plots and some PPCs), and may not fully characterise the uncertainty or potential biases associated with the posterior approximation for a single, specific observed dataset $\xobs$.

In this paper, we propose fully leveraging ensemble learning principles~\cite{Opitz_1999, polikar2006ensemble, rokach2010ensemble} to assess the quality of an SBI posterior approximation. 
Our core idea is conceptually extremely straightforward: if an SBI algorithm, trained on a finite set of simulations, were capable of reliably converging to the exact, true posterior distribution, then multiple instances of the same algorithm, trained independently on the same simulation budget or subsets thereof, should produce posterior estimates that agree perfectly with each other. 
Qualitatively, then, significant disagreement between an ensemble of trained estimators would signal potential issues such as high variance due to limited simulation data, poor convergence of the training process (e.g., neural network optimisation), instability in the approximation architecture itself, or data mismodelling when analysing real data.

The main idea of this work is to perform quality checks by computing and comparing discrepancy measures between the posterior estimates produced by each member of the ensemble. 
In particular, we highlight how the Kullback-Leibler (KL) divergence~\cite{kullback1951information} can be efficiently computed in an SBI context and provides a robust and interpretable quantity for comparison. 
This concept draws a clear analogy with diagnostic techniques used in Markov Chain Monte Carlo (MCMC) methods. 
In MCMC, multiple chains are often run in parallel from dispersed starting points. 
The convergence of these chains is assessed using diagnostics such as the Gelman-Rubin statistic ($\hat{R}$)~\cite{gelman1992inference}, which compares the variance within individual chains to the variance between chains. 
High $\hat{R}$ values track differences between the chains and indicate a lack of convergence, suggesting the samples do not yet represent the target posterior. 
Similarly, we propose below using the consistency (or lack thereof) across an ensemble of SBI estimators as one diagnostic for the reliability of the resulting posterior approximation.

We note that, of course, ensemble learning itself is a well-established paradigm in ML, where multiple individual models are trained and their predictions combined or compared to achieve better performance, robustness, and uncertainty quantification than any single model could provide (see, e.g.,~\cite{sagi2018ensemble} for a review). 
Common techniques in this regard include bagging, boosting, and stacking. 
By training an ensemble of SBI posterior estimators, we can not only assess convergence and stability but also potentially obtain more robust posterior estimates by combining the ensemble members, for instance, through mixture models or averaging, and gain insights into the approximation uncertainty. 
Ensemble methods have been explored before in the SBI context. 
For example, in Ref.~\cite{Alsing:2019xrx}, ensembles of likelihood estimators were used in an active learning framework to optimally generate new simulations with the acquisition protocol aimed at reducing likelihood uncertainty across the ensemble. 
In addition, Ref.~\cite{hermans2021trust} explored the coverage properties of ensembles, and Ref.~\cite{Cannon:2022aaa} demonstrated the behaviour of different SBI algorithms under various levels of model misspecification. 
More closely related to the current work, Ref.~\cite{Elsemuller:2023aaa} proposed the general principle of using deep ensemble disagreement to diagnose sensitivities arising from both insufficient training convergence and model misspecification. 
Our work provides a thorough validation and in-depth analysis of this diagnostic approach using a KL-divergence-based disagreement metric.

The rest of the paper is structured as follows. 
In Section~\ref{sec:theory} we establish the theoretical background for our approach and discuss the interpretation of the KL divergence in terms of systematic uncertainty estimation for the posteriors. 
Section~\ref{sec:KL_applied_to_ensemble} then demonstrates the practical application of these methods for assessing the convergence and reliability of SBI posteriors. 
We will demonstrate how ensembles can be used for generating a robust estimate of the systematic uncertainty in parameter reconstruction associated with the training procedure. 
Moreover, we show how the difference between the various networks in the ensemble can be used as a diagnostic tool to identify model misspecification. 
Finally, in Section~\ref{sec:conclusions}, we summarise our results and conclude.

\section{Ensemble Learning for Simulation-Based Inference}
\label{sec:theory}

\noindent Building upon the need for robust validation methods in SBI, this section develops the theoretical underpinnings of using ensemble methods in this context. 
We begin by very briefly reviewing the core concepts of ensemble learning within the broader context of machine learning. 
Subsequently, we discuss how these principles can be applied generally across different SBI methodologies. 
We then introduce a quantitative metric, the KL divergence matrix, designed to measure the consistency among different posterior estimators within an ensemble trained for a specific SBI task. 
This provides a means to assess the reliability of the inference results for a given observation. 
Finally, we explain how this metric can be used to place a lower bound on the systematic errors induced by the training and inference processes in SBI.

\subsection{Principles of Ensemble Learning}
\label{sec:ensemble_theory}

\noindent Ensemble learning constitutes a powerful paradigm in machine learning in which multiple individual models, often referred to as base learners or ensemble members, are constructed and their predictions are combined to produce a final result~\cite{dietterich2000ensemble, hastie2005elements}. 
The fundamental principle motivating this approach is that aggregating predictions from a diverse set of reasonably accurate models can lead to enhanced generalisation performance, increased robustness against noise or model misspecification, and more reliable uncertainty quantification compared to relying on any single constituent model.

Several techniques are common in ensemble learning to design the ensemble and aggregate the results. 
Bagging (or bootstrap aggregating) consists of training independent learners on different bootstrap samples (random samples drawn with replacement) of the original training dataset. 
Aggregating the predictions can lead to a reduced variance, which stabilises the predictions~\cite{breiman1996bagging}. 
Boosting consists of training a set of models sequentially, such that each subsequent model focuses on correcting the errors made by the models trained earlier in the sequence~\cite{freund1997decision, friedman2001greedy}. 
Boosting primarily aims to reduce the bias component of the error. 
Finally, stacking involves training multiple, often heterogeneous, base learners, with a separate meta-learner trained to optimally combine the predictions of these base learners~\cite{wolpert1992stacked}.

Beyond improving aggregate predictive accuracy, ensembles inherently provide a mechanism for assessing model stability and uncertainty. 
Disagreement among ensemble members trained under nominally similar conditions (for example, using the same algorithm and data budget but different random seeds or bootstrap samples) can denote high variance in the learning process or sensitivity to specific training instances~\cite{lakshminarayanan2017simple}. 
Although ensemble techniques have seen application within SBI, primarily aimed at improving the accuracy or robustness of the resulting posterior approximation~\cite{Alsing:2019xrx, hermans2021trust, yao2024simulation}, their systematic utilisation as a {\sl diagnostic} tool to gauge the reliability and convergence of the inference procedure for a {\sl single} specific observed dataset remains less explored. 
Addressing this problem is the central motivation for the present work.

\subsection{Simulation-Based Inference and Ensemble Applications}
\label{sec:sbi_ensembles}

\noindent SBI addresses the fundamental challenge of performing Bayesian inference when the likelihood function $p(\datavec|\thetavec)$, which quantifies the probability of observing data $\datavec$ given parameters $\thetavec$, is intractable or computationally prohibitive to evaluate directly/sample efficiently~\cite{cranmer2020frontier}.

SBI methods bypass this limitation by using a simulator, or forward model, capable of generating synthetic data $\datavec \sim p(\datavec|\thetavec)$ for any given parameter vector $\thetavec$ sampled from the prior distribution $p(\thetavec)$. 
The overarching goal of SBI methods is to approximate the posterior distribution $p(\thetavec|\xobs)$ for a specific, observed dataset $\xobs$. 
A variety of SBI algorithms/strategies have been developed for this purpose, including: Neural Posterior Estimation (NPE), which directly targets the posterior distribution; Neural Likelihood Estimation (NLE), which instead estimates the likelihood and subsequently samples it; and Neural Ratio Estimation (NRE), which finally constructs a neural estimator of the likelihood-to-evidence ratio that can also be sampled. 
Despite these diverse mechanisms, the unifying feature of modern SBI methods is the training of a statistical model using a finite set of simulations to approximate the posterior $q_\phi(\theta | x)$ or a related quantity. 
The principles of ensemble learning from Section \ref{sec:ensemble_theory} can be naturally applied in this context. 
Instead of training a single estimator $q_\phi(\thetavec|\datavec)$, we train an ensemble of $M$ posterior estimators, $\{q_{m}(\thetavec|\datavec)\}_{m=1}^M$. 
Each of these can be considered as an approximation to the target posterior for the given observation $\xobs$.

Our central hypothesis is that the {\sl consistency} among these approximations can serve as a valuable diagnostic. 
If the SBI training process is stable and has converged reliably for observation $\xobs$, the posterior estimates $q_m(\thetavec|\xobs)$ from different ensemble members should be measureably similar. 
Conversely, significant divergence among them signals potential problems, such as insufficient simulation budget, poor optimiser convergence, or inherent difficulties in approximating the posterior for that specific $\xobs$ (e.g., due to mismodelling).

\subsection{The KL Divergence Matrix for Ensemble Assessment}
\label{sec:kl_metric}

\noindent To quantify the (dis)agreement between these posterior approximations, we employ the Kullback-Leibler (KL) divergence. 
The KL divergence between a probability distribution function (PDF) $Q$ and another PDF $P$ measures the information lost when $Q$ is used to approximate $P$. 
It is defined as:
\begin{equation}
\label{eq:kl_definition}
\KL{P}{Q} = \int P(\thetavec) \log \frac{P(\thetavec)}{Q(\thetavec)} \, \mathrm{d}\thetavec.
\end{equation}
The measure $\KL{P}{Q} \ge 0$ with equality being approached if and only if $P=Q$ almost everywhere. 
From the definition, we also see that the KL is in general asymmetric, i.e., $\KL{P}{Q} \neq \KL{Q}{P}$ for arbitrary $P$ and $Q$. 
In our context, we will compute the KL divergence between pairs of posterior approximations $q_i(\thetavec|\xobs)$ and $q_j(\thetavec|\xobs)$ from the ensemble.

To systematically capture pairwise disagreements within the ensemble of posteriors $\{q_m(\thetavec|\xobs)\}_{m=1}^M$ for a specific observation $\xobs$, we propose to construct an $M \times M$ matrix, termed the ``KL matrix", $\mathbf{K}$. 
The elements of this matrix are defined to be the KL divergences between the respective posterior approximations:
\begin{equation}
\label{eq:kl_matrix}
K_{ij} = \KL{q_i(\thetavec|\xobs)}{q_j(\thetavec|\xobs)}.
\end{equation}
By definition, the diagonal elements $K_{ii}$ are zero. 
The off-diagonal elements $K_{ij}$ ($i \neq j$) quantify the divergence from $q_j$ to $q_i$, representing the information lost when approximating $q_i$ using $q_j$.

A practical advantage, particularly relevant for SBI methods based on neural density estimators for the posterior (such as NPE), is the feasibility of estimating these KL divergences using Monte Carlo sampling. 
Specifically, $K_{ij}$ can be approximated by drawing $N_s$ samples\footnote{To be more precise, enough samples should be drawn such that the Monte Carlo error in Eq.~\eqref{eq:kl_monte_carlo} falls below a chosen threshold.} $\thetavec^{(k)}$ from the distribution $q_i(\thetavec|\xobs)$ and averaging the log-ratio of the probability densities:
\begin{equation}
\label{eq:kl_monte_carlo}
K_{ij} \approx \frac{1}{N_s} \sum_{k=1}^{N_s} \left[ \log q_i(\thetavec^{(k)}|\xobs) - \log q_j(\thetavec^{(k)}|\xobs) \right].
\end{equation}
All this estimation requires is the ability to both sample from each $q_m(\thetavec|\xobs)$ and evaluate the log-densities $\log q_m(\thetavec|\xobs)$, capabilities readily available in many modern SBI implementations. 
While generating samples is simple and efficient in the NPE case, it can be more costly for alternative approaches like NRE or NLE. 
However, this does not mean that estimating the components $K_{ij}$ is also computationally expensive. 
For NRE, each ensemble member learns the ratio $r_i(\theta, x) \approx p(x|\theta)/p(x)$, leading to a posterior estimate $q_i(\theta|x_0) \propto r_i(\theta, x_0)p(\theta)$. 
The log-posterior difference needed for the KL calculation in Eq.~\eqref{eq:kl_monte_carlo} simplifies to $\log r_i(\thetavec^{(k)},\xobs) - \log r_j(\thetavec^{(k)}, \xobs)$, as the prior $p(\theta)$ cancels. 
As such, we see that the primary cost for NRE is generating samples from each ensemble member with stochastic sampling, after which computing the KL matrix only requires evaluating the learned ratio. 
For NLE, each member learns the likelihood $\tilde{q}_i(x|\theta) \approx p(x|\theta)$. This means that the posterior approximation is given by $q_i(\theta | x) = \tilde{q}_i(x | \theta)p(\theta) / \tilde{p}_i(x)$, where $\tilde{p}_i(x)$ is the approximate evidence of the data $x$ for each ensemble member $i$. In this case, the log-posterior difference involves both the log-likelihood difference and the log of the Bayes factor between members. 
To compute this in full, this requires \emph{both} posterior samples from each estimator $\tilde{q}_i(\xobs | \theta)$ and an estimate of the evidence for each member $\tilde{p}_i(\xobs)$. In general, this latter component may be challenging, however, it can be computed along with posterior sampling using techniques like nested sampling. 
Thus, while the KL framework is most efficiently applied in NPE, it remains conceptually and practically applicable to NRE and NLE, with the main cost being the stochastic sampling step already required to obtain posterior constraints.

The KL matrix $\mathbf{K}$ serves as a rich diagnostic tool, allowing for quantitative assessment of ensemble consistency:
\begin{itemize}
    \item \textbf{Magnitude of Off-Diagonal Elements:} Small values across all $K_{ij}$ ($i \neq j$) indicate high consistency; the ensemble members have converged to similar posterior approximations. 
    This suggests the inference process was stable for $\xobs$. 
    Conversely, large values signal significant disagreement, potentially arising from insufficient training data (simulations), poor convergence of the underlying optimisation for some members, multimodality in the optimisation landscape, or inherent instability in approximating the posterior at $\xobs$.
    \item \textbf{Structure and Asymmetry:} Examining the matrix structure can reveal patterns. 
    For instance, if one estimator $q_k$ consistently yields large $K_{ik}$ or $K_{kj}$ values when paired with others, it might be an outlier. 
    Asymmetry ($K_{ij} \neq K_{ji}$) can indicate differences in the distributional shapes; for example, if $q_j$ is significantly more diffuse than $q_i$, $K_{ij}$ (approximating the broader $q_j$ with the narrower $q_i$) might be substantially larger than $K_{ji}$ (approximating the narrower $q_i$ with the broader $q_j$).
\end{itemize}
Crucially, the KL matrix provides a way to benchmark the internal consistency of the ensemble, which acts as a proxy for a \emph{lower bound} on the systematic error associated with the inference process. 
A large value in the KL matrix, specifically a large $\max_{i \neq j} K_{ij}$, indicates significant disagreement between at least two ensemble members. 
Since the true posterior $p(\thetavec|\xobs)$ is unique (assuming a well-defined problem), if two approximations $q_i$ and $q_j$ diverge significantly from each other (i.e., $K_{ij}$ or $K_{ji}$ is large), they cannot both be arbitrarily close to the true posterior. 
Thus, a large $\max_{i \neq j} K_{ij}$ serves as a quantitative indicator of ensemble instability or unreliability for the given observation $\xobs$. 
It suggests that the inference procedure yields significantly different outcomes under slight perturbations (such as different initialisations or data samples), casting doubt on the trustworthiness of any single member or even an aggregated result as a faithful approximation of the true posterior. 
While low KL values do not guarantee accuracy (all members could converge to the same wrong answer), high KL values strongly signal a problem with the consistency of the inference for that specific $\xobs$. 
This quantitative diagnostic forms the basis for the tools and analyses presented in the following sections.

\subsection{Interpreting KL divergence as systematic uncertainty}
\label{sec:kl_gaussian}

\begin{figure*}[ht]
    \centering
\includegraphics[width=\linewidth]{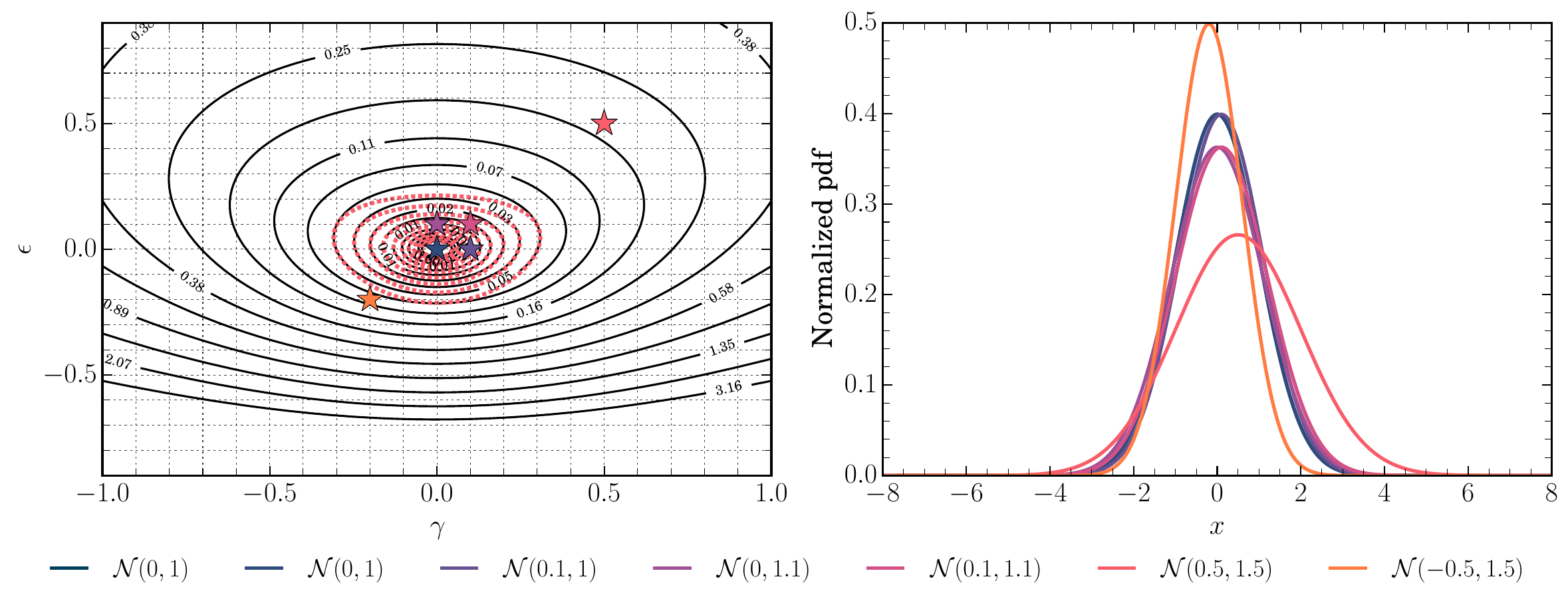}
    \caption{\textbf{Interpreting the KL divergence:} A set of examples illustrating quantitatively how the KL divergence depends on the differences in the means/variances of 1-d Normal distributions. 
    Here, the base distribution $\mathcal{N}_1$ has $\mu_1=0$ and $\sigma_1=1$ such that $\mu_2 = \gamma$ and $\sigma_2=1+\epsilon$ are the parameters of the contrasting distribution $\mathcal{N}_2$. 
    \textit{Left:} Contour plot showing the value of the KL divergence for different values of $\epsilon$ and $\gamma$. 
    Also shown (in red) is the quadratic approximation derived in Eq.~\eqref{eq:KL_approx}, which we see matches well for small divergences. 
    The coloured stars correspond to specific values reported in the legend. 
    \emph{Right:} 1d histograms showing the PDFs for the values of $\epsilon$ and $\gamma$ corresponding to the stars in the left plot.}
    \label{fig:KL_2d}
\end{figure*}

\noindent A key aspect of our framework is interpreting the KL divergence as a lower bound on the systematic uncertainty of the posterior approximation. 
To understand what a given KL value implies about the disagreement between posteriors, it is helpful to consider the case of two one-dimensional, Gaussian distributions. 
For this case, it is simple to derive an analytical expression for the KL divergence, 
\begin{equation}
    \KL{\mathcal{N}_1}{\mathcal{N}_2} = \log\left( \frac{\sigma_2}{\sigma_1}\right) + \frac{\sigma_1^2 + (\mu_1 - \mu_2)^2}{2 \sigma_2^2} - \frac{1}{2}\,,
    \label{eq:KL_1d_gaussian}
\end{equation}
where $\mu_1$, $\mu_2$, $\sigma_1$, and $\sigma_2$ are the means and standard deviations of the two distributions, respectively. 
Whether a given level of KL divergence is "too large" depends entirely on the application context.
If an analysis requires parameter constraints more precise than this systematic uncertainty, then such a KL value indicates that the ensemble has not achieved sufficient consistency. 
The translation given in this section is most accurate for approximately Gaussian posteriors; for highly non-Gaussian or multimodal distributions, the KL value serves as a general inconsistency metric which can be benchmarked across the ensemble.

To better understand the mapping between the KL divergence and $\sigma$-levels, let us proceed further by considering small perturbations between the two distributions. 
We parametrize this as $\sigma_2=\sigma_1(1+\epsilon)$ and $\mu_2 = \mu_1+\gamma \sigma_1$ such that $\epsilon$, $\gamma \ll 1$ are perturbations in units of $\sigma_1$. 
With this parameterisation, we can obtain the perturbative result,
\begin{equation}\label{eq:KL_approx}
    \KL{\mathcal{N}_1}{\mathcal{N}_2} \simeq  \frac{\gamma^2}{2} + \epsilon^2 - \gamma^2\epsilon\; ,
\end{equation}
at the third order in the mean and standard deviation differences. 
An example of the KL divergence dependence on $\gamma$ and $\epsilon$ is shown in Fig.~\ref{fig:KL_2d} for a simple case of two 1-d Gaussian distributions. 
From Eq.~\eqref{eq:KL_approx}, we see that the KL divergence scales quadratically for small differences in both the mean and widths of the distributions. 
With large differences, however, the divergence becomes more complex and can induce correlated effects. 
This suggests that the KL divergence has a simple interpretation when the differences between distributions are small, which is the regime we aim to work in when training SBI models. 
For reference, in Fig.~\ref{fig:KL_2d}, we show a set of benchmark examples to provide some graphical intuition regarding typical numerical values of the KL divergence.

The generalization of Eq.~\eqref{eq:KL_1d_gaussian} to $n$-dimensional Gaussian distributions is
\begin{equation}
\begin{aligned}
\KL{\mathcal{N}_1}{\mathcal{N}_2} & = \frac{1}{2} \left[(\vec{\mu}_2  - \vec{\mu}_1)^\dagger \Sigma_2^{-1} (\vec{\mu}_2 - \vec{\mu}_1)  \right. 
\\ & + \left. 
    \log \frac{|\Sigma_2|}{|\Sigma_1|} +  {\rm Tr}(\Sigma_2^{-1} \Sigma_1) - n \right]\,,
\end{aligned}
\end{equation}
where $\vec{\mu}_{1,2}$ and $\Sigma_{1,2}$ are the mean vectors and covariance matrices respectively. 
In this case, we can gain insight into the perturbative limit using a similar approach to the one-dimensional case. 
We can diagonalise the system about the eigenvector basis $U$ of $\Sigma_1$ by assuming $\Sigma_1 = U\Lambda^2 U^\dagger$ with $\Lambda = \mbox{diag}(\sigma_1,...,\sigma_n)$ and defining perturbations $\Sigma_2 = U\Lambda^{\prime 2} U^\dagger$ and $U^\dagger(\vec{\mu}_2-\vec{\mu}_1) = \Gamma \Lambda \vec{1}$ where $\Lambda^{\prime} = \Lambda(\mathbb{I}+\mathcal{E})$,  $\mathcal{E} = \mbox{diag}(\epsilon_1,...,\epsilon_n)$, and $\Gamma = \mbox{diag}(\gamma_1,...,\gamma_n)$ . 
Then, in the diagonal basis, we recover
\begin{equation}
    \KL{\mathcal{N}_1}{\mathcal{N}_2} \simeq  \sum_i^n\frac{\gamma_i^2}{2} + \epsilon_i^2 - \gamma_i^2\epsilon_i\;,
\end{equation}
which is indicative of how the KL divergence measure scales with dimensionality in the perturbative Gaussian limit. 
From this equation, it is manifest that the mismatch is expected to scale linearly with the number of dimensions. 
This can be useful to gauge the mismatch in higher-dimensional cases by comparing to the limiting perturbative Gaussian case. 
Guided by these considerations, we propose $\max_{ij}(\textbf{K}/n) \ll 1$ as a heuristic necessary condition for consistency between models in the SBI context.

\begin{figure*}[t]
    \centering
\includegraphics[width=\linewidth]{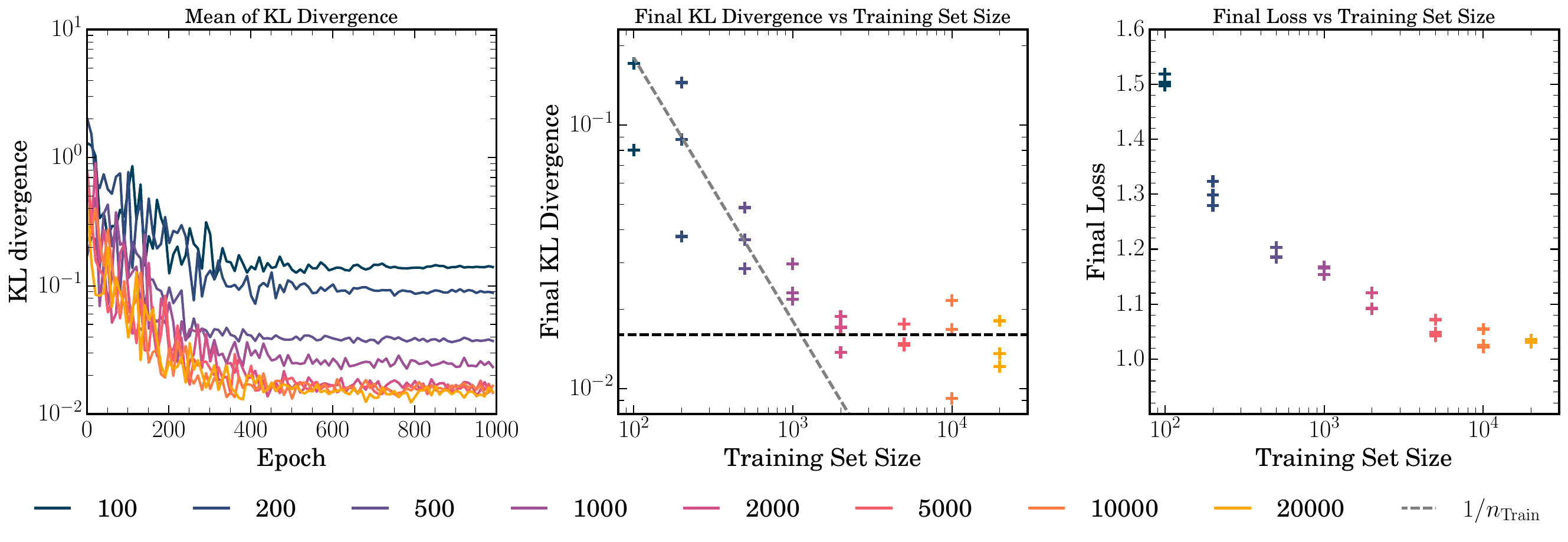}
    \caption{\label{fig:KL_GM_all} \textbf{KL matrix as a training diagnostic.} \emph{Left:} The mean pairwise KL divergence, monitored during the training of single ensembles. 
    Each curve represents one ensemble trained with a specific dataset size (see legend), demonstrating how the KL matrix can be used to track convergence and determine when further training is unlikely to improve consistency. 
    \emph{Middle:} Value of the KL divergences in the ensemble (averaged over the last 30 epochs) at the end of the training as a function of training dataset size. 
    In dashed grey, we show the expected $1/n_{\rm Train}$ behaviour before saturating the intrinsic variance floor. 
    \emph{Right:} Value of the loss function at the end of the training process as a function of the training dataset size.}
\end{figure*}

Another consideration is how the KL divergence is expected to scale as the size of the training set varies. 
In general, for an infinitely complex network capable of perfectly approximating the posterior, trained with a perfect strategy, we expect the divergence to endlessly scale with the inverse of the training set size (i.e., with more and more data, we expect the approximation to become better and better). 
However, for finite networks, an intrinsic variance due to imperfect training/network architecture will eventually set a floor (see the discussion around Fig.~\ref{fig:KL_GM_all}). 
To better understand the scaling behaviour of KL with the size of the training dataset $n_{\rm Train}$, we can consider two models $\hat{P}_1(\theta)$ and $\hat{P}_2(\theta)$ trained to estimate the true PDF $P(\theta)$. 
We can assume that the offsets between estimators are small, such that $\hat{P}_i(\theta) = P(\theta) + \delta P_i (\theta)$, with $\delta P_i (\theta)$ denoting the residual with respect to $P(\theta)$. 
Using this setup, we can express the KL divergence between the two estimators as:
\begin{equation}
\begin{aligned}
\label{eq:kl_diff}
\KL{\hat{P}_1 }{ \hat{P}_2} 
& \simeq \int \left[  \delta P_1(\thetavec) - \delta P_2(\thetavec)  + \right.
\\ 
& \hspace{-1cm} \left. 
    \hspace{.2cm} + \frac{ \delta P_1^2(\thetavec) +\delta P_2^2(\thetavec) -2 \delta P_1(\thetavec) \delta P_2(\thetavec)  }{P(\thetavec)} \right] \, \textrm{d}\thetavec .
\end{aligned}
\end{equation}
We can proceed further by assuming the estimators to be unbiased, implying the expectation value (over the stochasticity due to initial random weights and steps involved in the training process) of the two models is $P(\theta)$, and $\langle \delta P_1(\thetavec) \rangle = \langle \delta P_2(\thetavec) \rangle = 0$. 
Then, the expectation value of Eq.~\eqref{eq:kl_diff} simplifies to
\begin{equation}
\label{eq:KL_expectation}
  \langle \KL{\hat{P}_1 }{ \hat{P}_2} \rangle  \simeq \int \frac{ \langle\delta P_1^2(\thetavec)\rangle +\langle \delta P_2^2(\thetavec) \rangle }{P(\thetavec)}  \, \textrm{d}\thetavec \propto \frac{1}{n_{\rm Train}} \; ,
\end{equation}
where we have assumed that, since $\hat{P}_i (\theta) $ are estimators of the true PDF, we can expect that, in the best-case scenario, the accuracy (i.e., the standard deviation) of such an estimate scales as $1/\sqrt{ n_{\rm Train}}$. 
In such a case, the expectation value of $\KL{\hat{P}_1 }{ \hat{P}_2}$ would scale as $1 / n_{\rm Train}$. 
In the following sections, we will present an explicit example where we find good agreement with this analytical estimate.

\section{KL divergence applied to ensemble training}
\label{sec:KL_applied_to_ensemble}

\noindent In this section, we restrict ourselves to the case of Neural Posterior Estimation (NPE) and explore some applications of the KL matrix as a diagnostic tool for SBI.\footnote{For simplicity, for all the examples considered in this work, we have only considered the upper triangular part of the $\textbf{K}$ matrix, as $K_{ij}$ and $K_{ji}$ provide correlated information. 
The asymmetry ($K_{ij} \neq K_{ji}$) can provide additional diagnostic information about the relative shapes of the posteriors, but we do not explore it in this work as we found it to be a subdominant effect in our experiments.}
In Section~\ref{sec:NPE_basics}, we begin by reviewing the main ideas behind NPE. 
Then, in Section~\ref{sec:NPE_KL_convergence}, we discuss the use of the KL matrix to assess the convergence among different networks in an ensemble. 

Finally, in Section~\ref{sec:NPE_KL_misspecification}, we explore the usage of the KL matrix to identify model misspecification.

\subsection{Basics of Neural Posterior Estimation (NPE)}
\label{sec:NPE_basics}
\noindent As discussed in Section~\ref{sec:sbi_ensembles}, the core idea of NPE is to approximate the posterior distribution $p(\theta|x)$ using a neural density estimator, such as a normalising flow. 
A standard choice of loss function to train this type of model is
\begin{equation}
\begin{aligned}
    \mathcal{L}(\phi) & = - \mathbb{E}_{p(\thetavec, \datavec)} [\log q_\phi(\thetavec|\datavec)] \\
    & = - \int p(\thetavec, \datavec) \log q_\phi(\thetavec|\datavec) \, \mathrm{d}\thetavec \, \mathrm{d}\datavec,
\end{aligned}
\label{eq:npe_loss}
\end{equation}
where $\phi$ are the parameters of the network and $q_{\phi}(\theta|x)$ is the approximate trained posterior. 
In practice, the integral in the second line of Eq.~\eqref{eq:npe_loss} is approximated using a finite dataset of $N$ simulated pairs $(\theta_i, x_i) \sim p(\theta, x)$ via a Monte Carlo average: 
\begin{equation}
    \mathcal{L}(\phi) \approx - \frac{1}{N} \sum_{i=1}^N \log q_\phi(\thetavec_i | \datavec_i) \; .
\end{equation}
It is easy to show that, up to factors that do not depend on the network parameters, such a loss function is proportional to the KL divergence between the true posterior $p(\thetavec|\datavec)$ and the learned approximation $q_\phi(\thetavec|\datavec)$, averaged over the marginal data distribution $p(\datavec) = \int p(\datavec|\thetavec)p(\thetavec) \mathrm{d}\thetavec$:
\begin{equation}
\begin{aligned}
&\mathcal{L}(\phi) =
- \int p(\datavec) p(\thetavec|\datavec) \log q_\phi(\thetavec|\datavec) \, \mathrm{d}\thetavec \, \mathrm{d}\datavec \\
&= 
\int p(\datavec) \left( \int p(\thetavec|\datavec) \log \frac{p(\thetavec|\datavec)}{q_\phi(\thetavec|\datavec)} \, \mathrm{d}\thetavec \right) \mathrm{d}\datavec + \mathrm{const.} \\
&= 
\mathbb{E}_{p(\datavec)} \left[\KL{p(\thetavec|\datavec)}{q_\phi(\thetavec|\datavec)}\right] + \mathrm{const.},
\label{eq:npe_kl_connection}
\end{aligned}
\end{equation}
so that minimising the loss function is equivalent to minimising the KL divergence between $p(\thetavec|\datavec)$ and $q_{\phi}(\thetavec|\datavec)$. 
This connection highlights that NPE seeks an approximation $q_\phi$ that is close to the true posterior $p(\thetavec|\datavec)$ in an information-theoretic sense, on average across possible datasets $\datavec$ drawn from the model. 
Moreover, it gives additional motivation for our choice of the KL divergence as a diagnostic tool for SBI. 
In particular, given two learned approximants $q_{i}(\thetavec|\datavec)$ and $q_{j}(\thetavec|\datavec)$ of $p(\theta|x)$, it is possible to construct counterexamples where the KL divergences between $p$ and $q_{i}$, as well as $p$ and $q_{j}$ are small, while the KL divergence between $q_{i}$, $q_{j}$ is larger. 
Thus, computing the KL divergence matrix for the networks in the ensemble defines a test that provides \emph{independent} information compared to the loss function.

As discussed above, a particular advantage of NPE in this context is the relative ease of sampling and evaluating the log-density of the approximate posterior. 
This makes estimating the KL divergences across ensembles especially fast and efficient. 
However, we emphasise that there are no fundamental barriers to evaluating KL for the other SBI paradigms mentioned in Section~\ref{sec:sbi_ensembles}.

\subsection{Assessing ensemble convergence and estimating systematic errors}
\label{sec:NPE_KL_convergence}

\begin{figure*}[ht]
    \centering
\includegraphics[width=.24\linewidth]{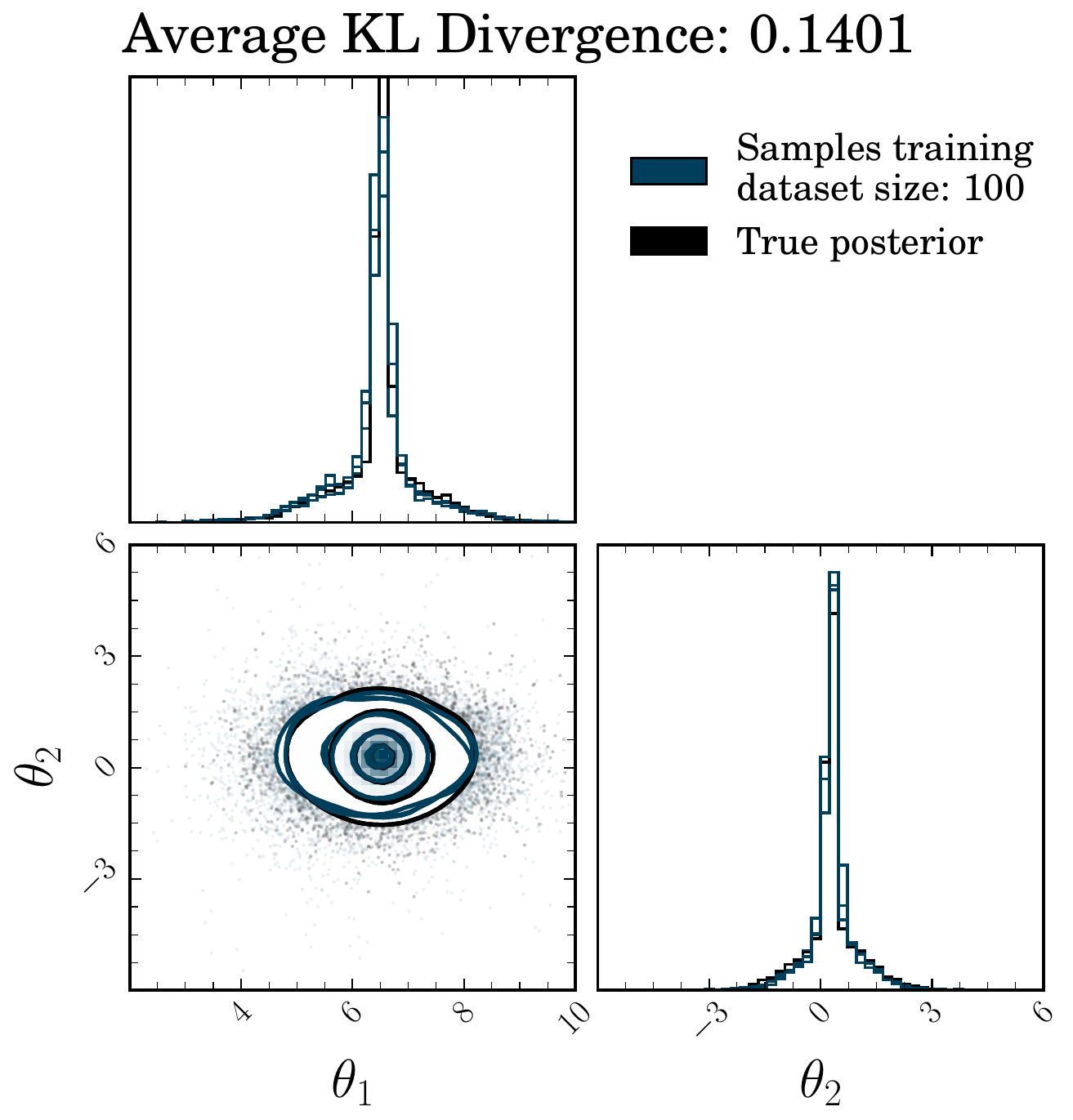}
\includegraphics[width=.24\linewidth]{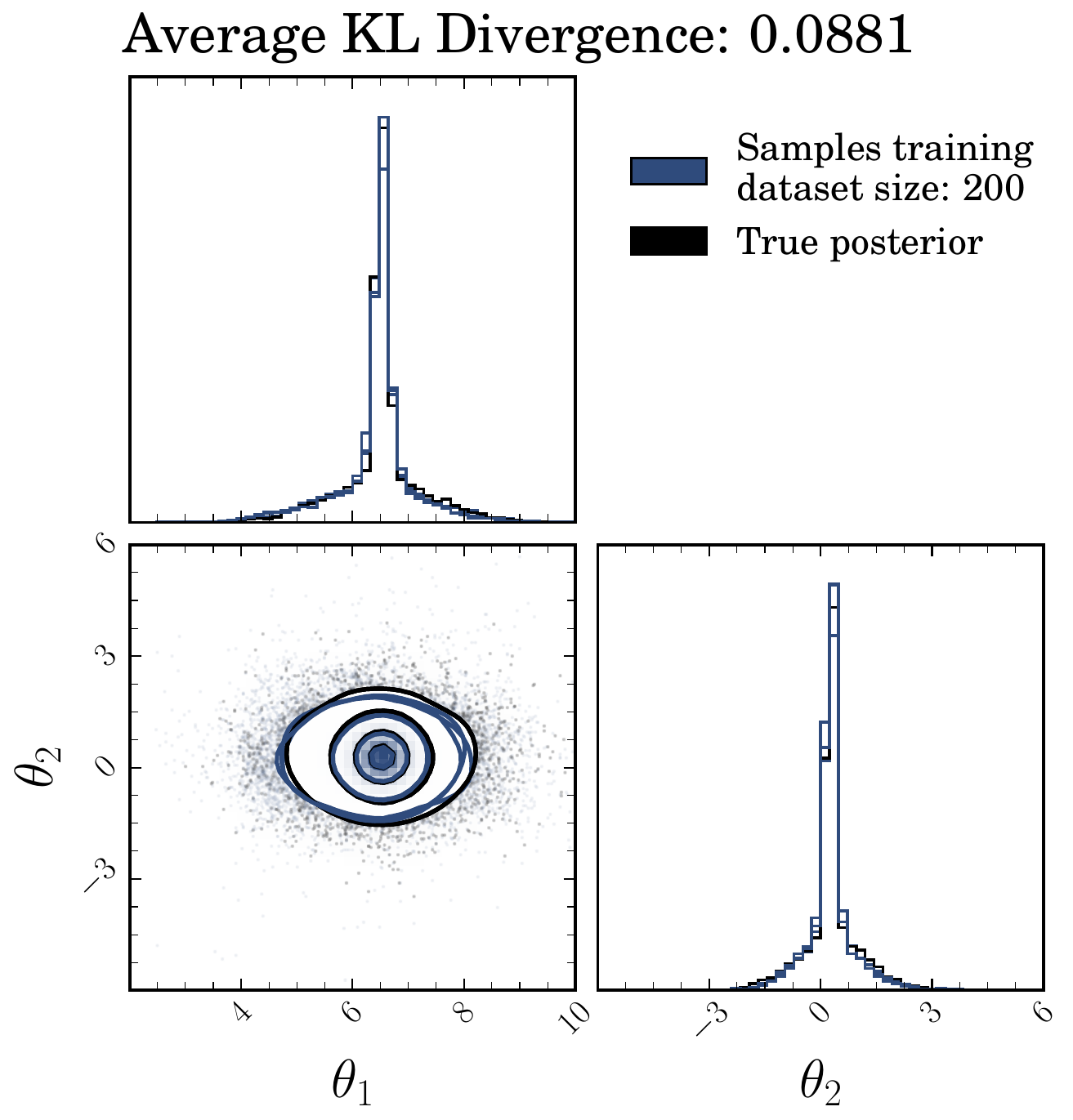}
\includegraphics[width=.24\linewidth]{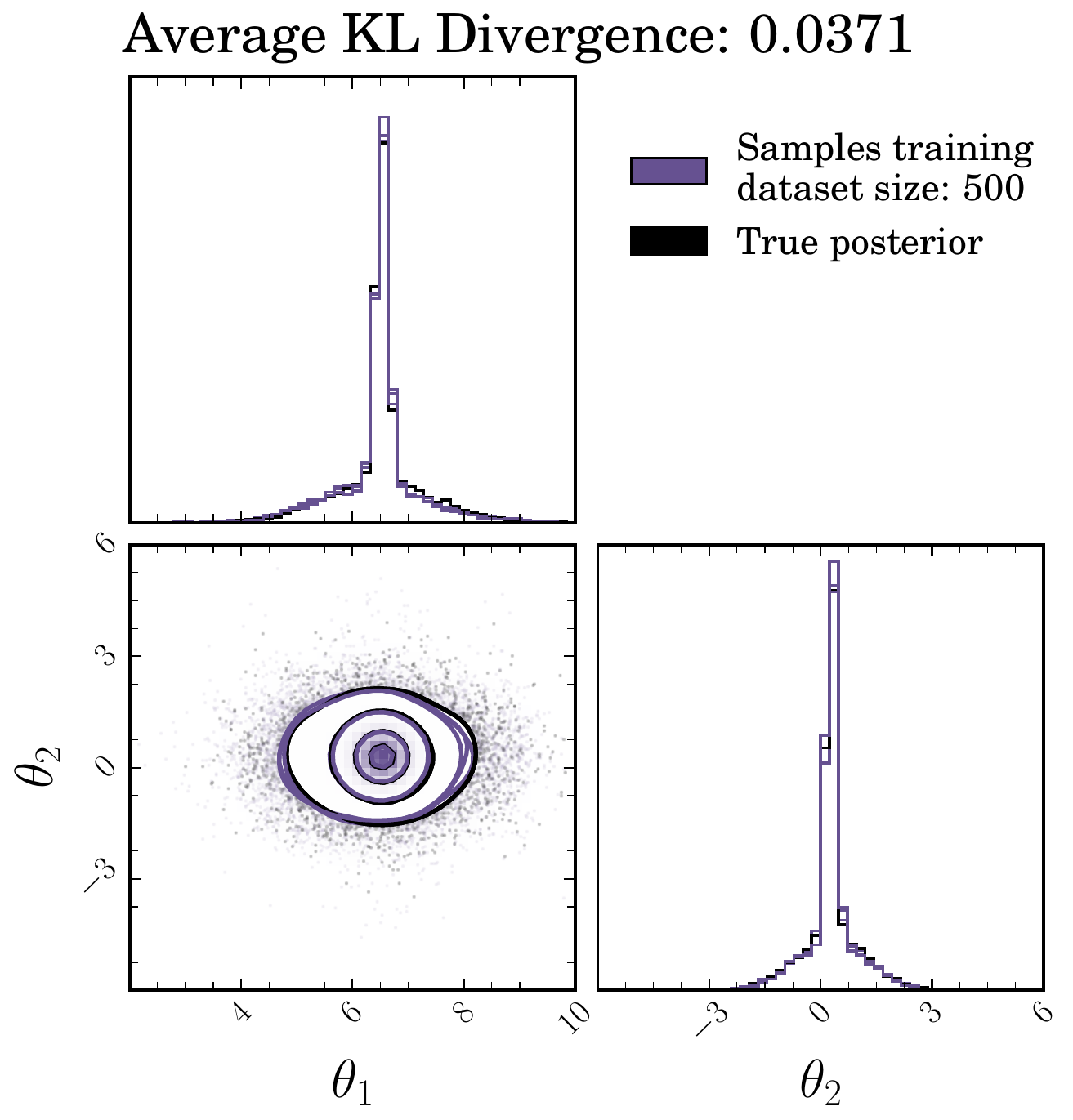}
\includegraphics[width=.24\linewidth]{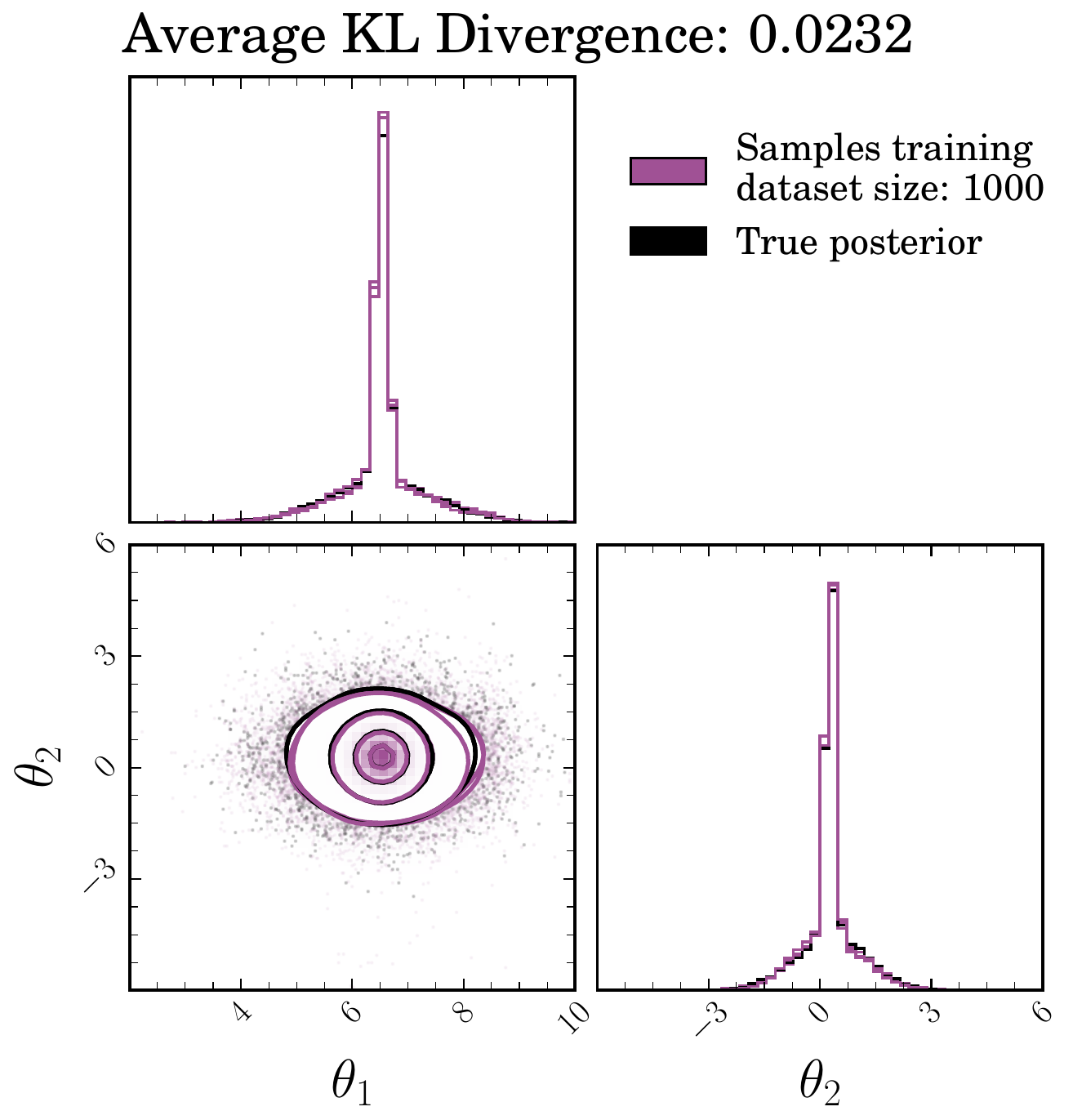}\\
\includegraphics[width=.24\linewidth]{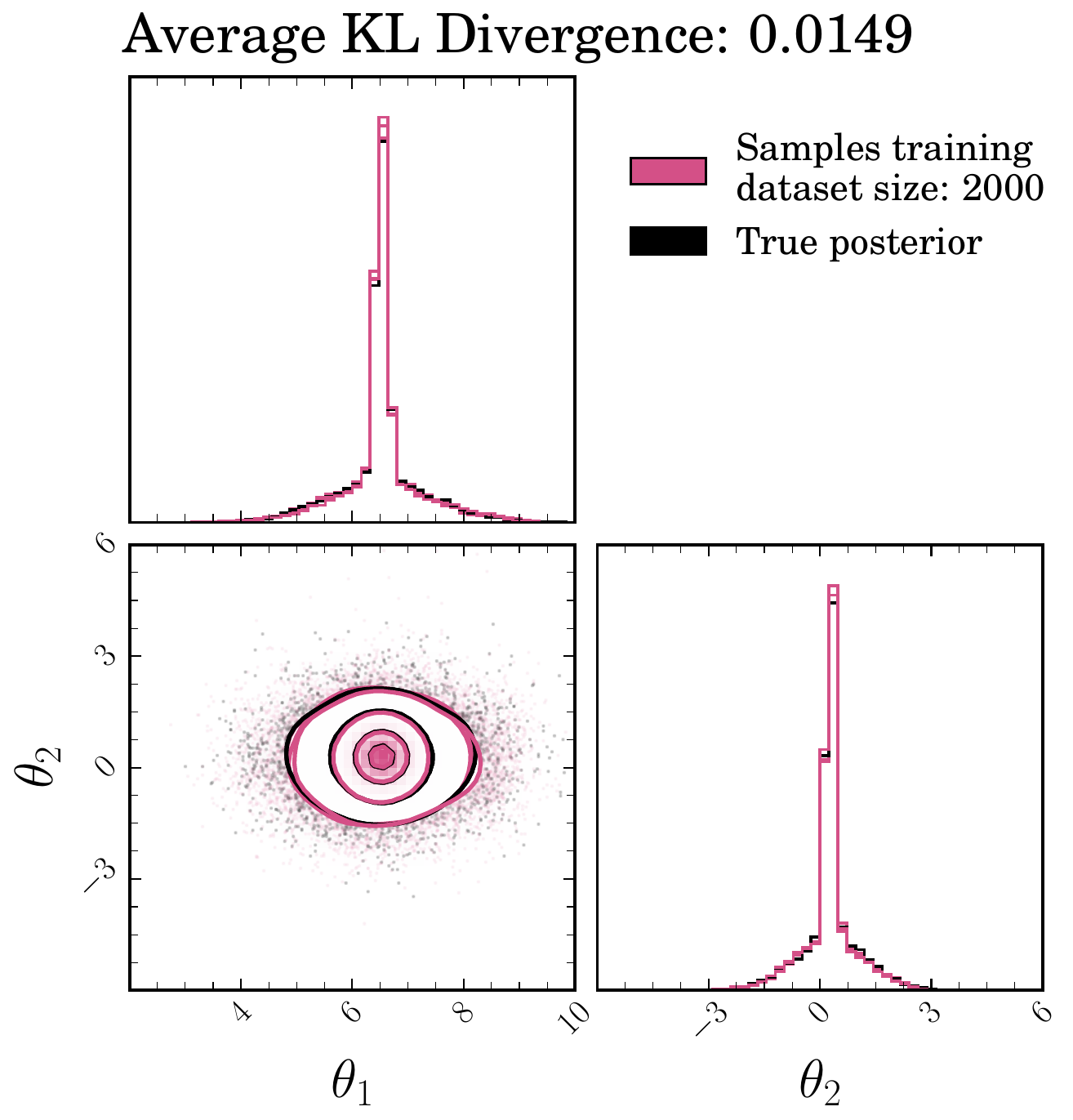}
\includegraphics[width=.24\linewidth]{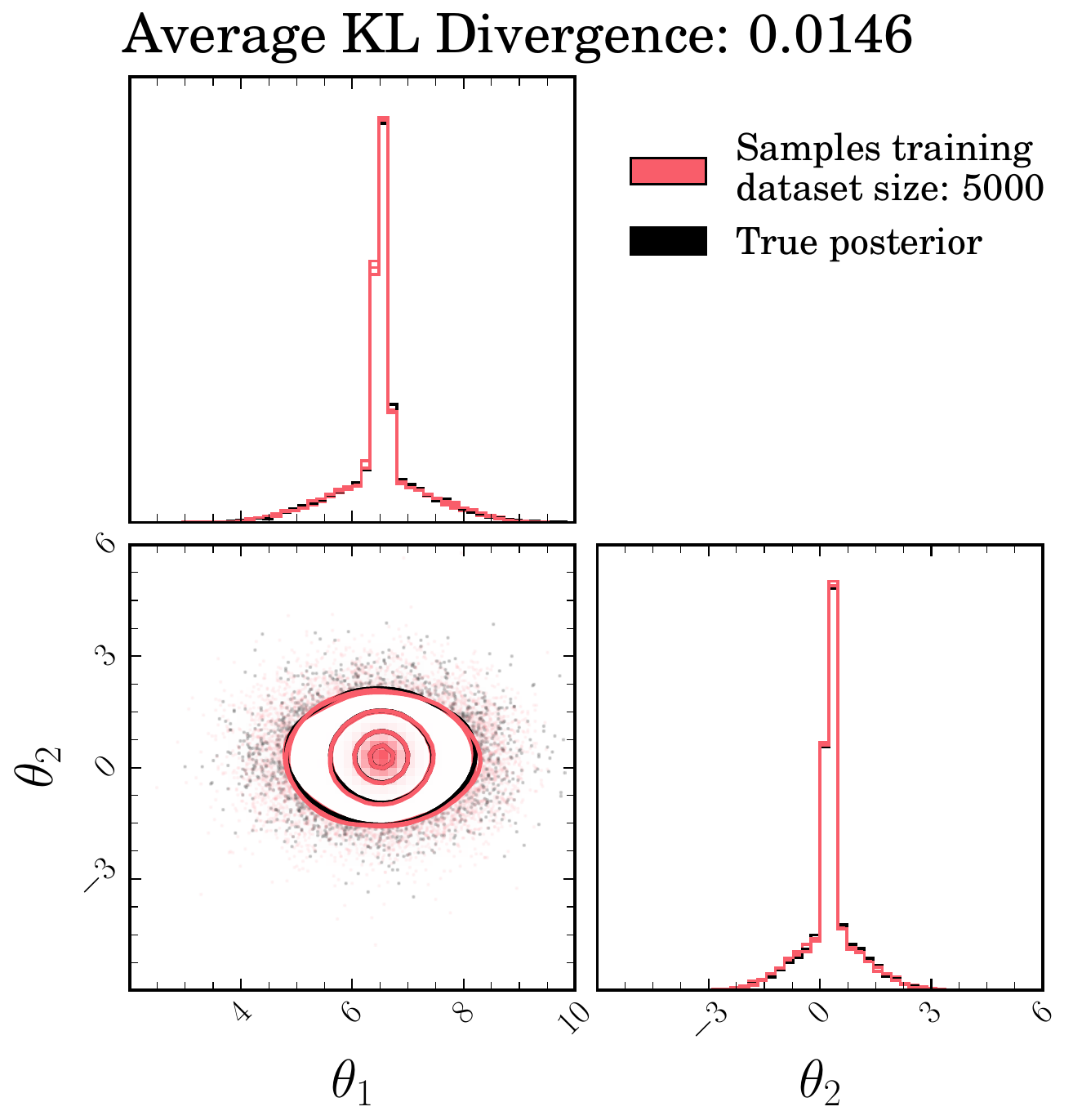}
\includegraphics[width=.24\linewidth]{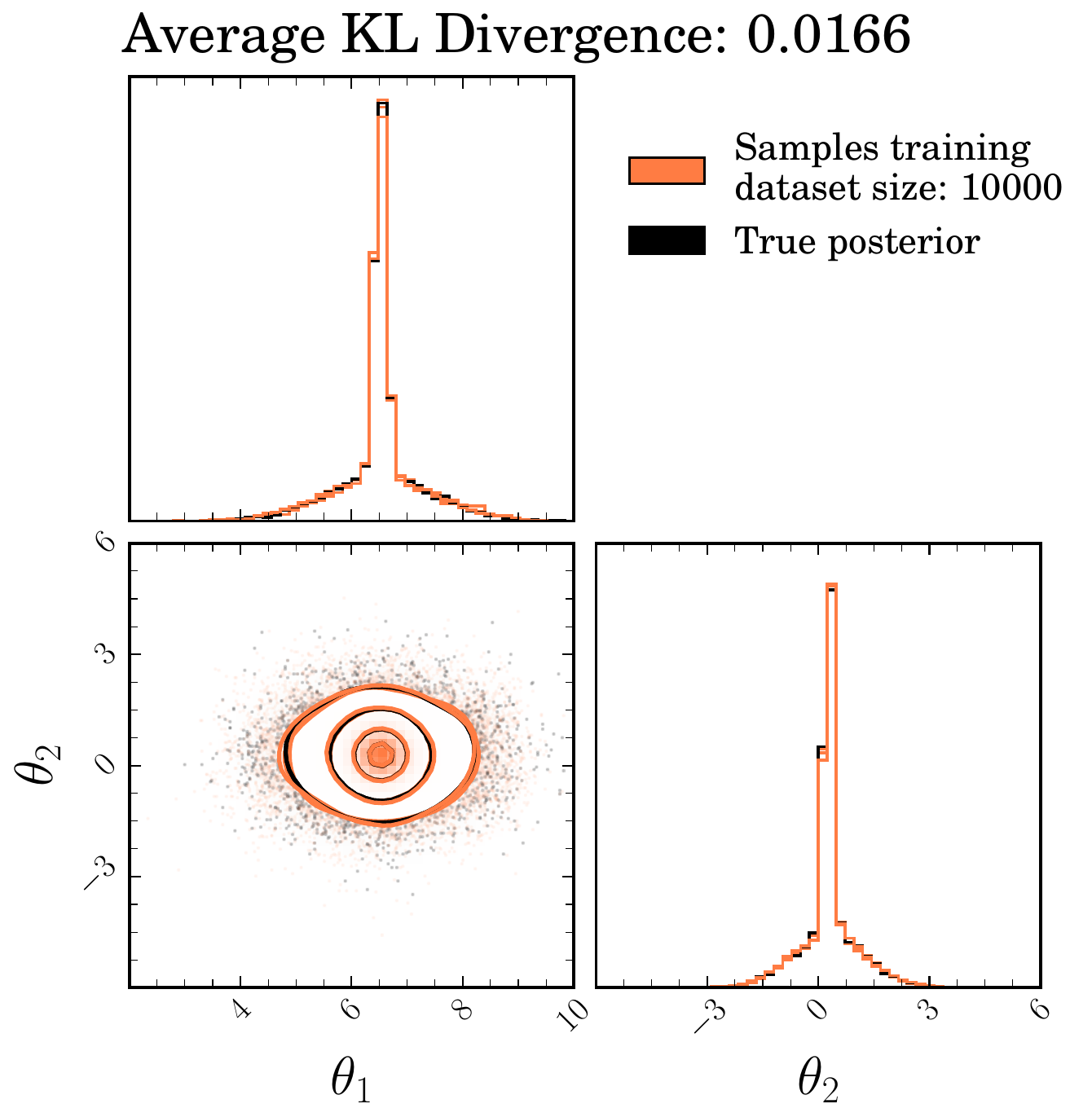}
\includegraphics[width=.24\linewidth]{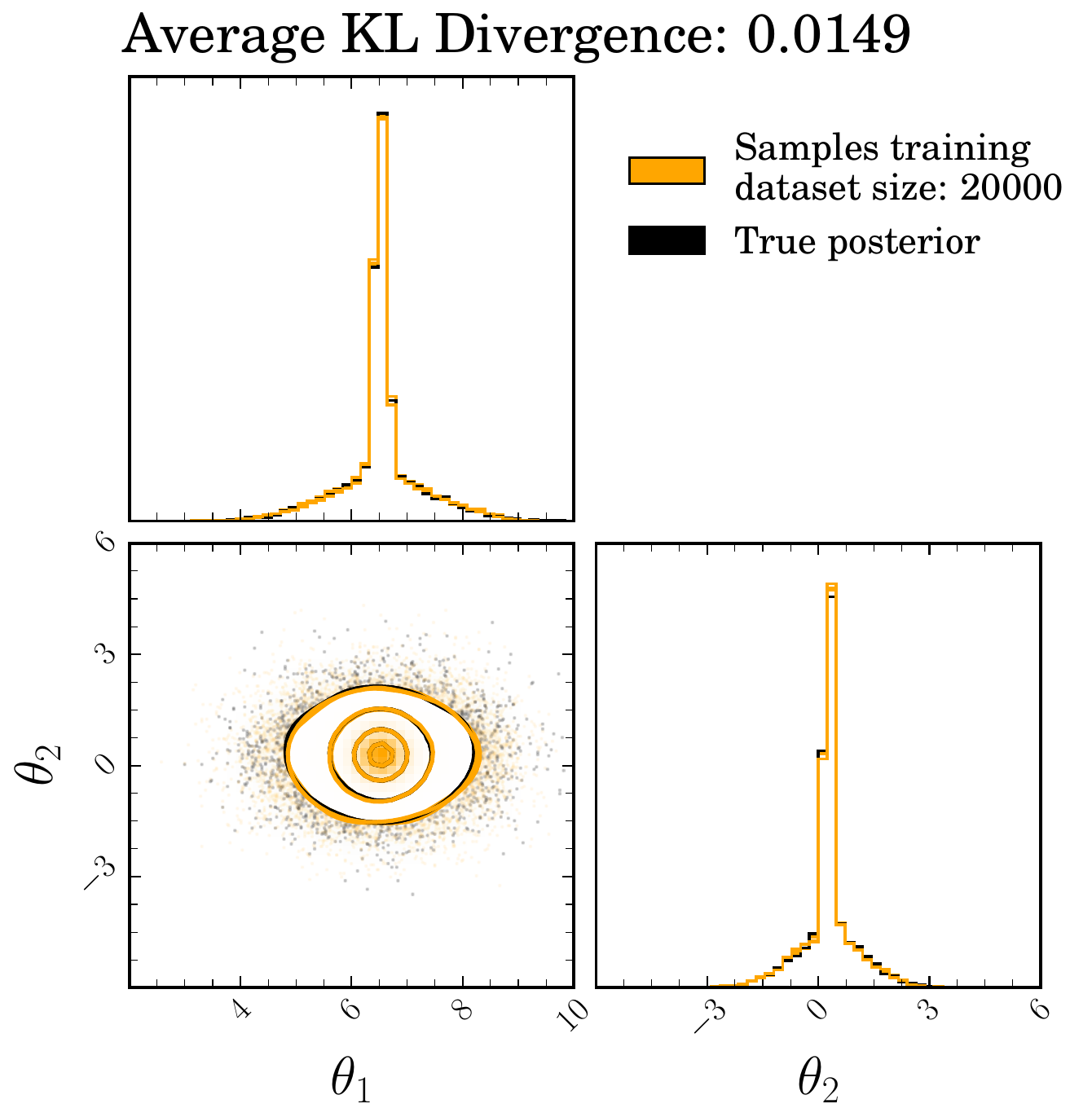}
\caption{\label{fig:corners} \textbf{Evolution of the posteriors.} True posterior distribution (black) compared to samples drawn from the different networks in the ensembles. 
The different panels (top left to bottom right) correspond to different training dataset sizes (colour code as in Fig.~\ref{fig:KL_GM_all}). 
Each panel shows the results from a single ensemble trained with a specific dataset size; multiple ensembles are only needed to compare how systematic uncertainties vary with training conditions, not to estimate the uncertainty for a single setup.}
\end{figure*}

\noindent For the first illustrative example demonstrating the application of the KL matrix as a diagnostic test, we consider training an ensemble of identical networks with initial random weights for a specific inference task. 
For this purpose, we choose a simple case study where the simulator corresponds to a two-dimensional Gaussian mixture model from the \texttt{sbibm} library~\cite{lueckmann2021benchmarking}. 
In this context we can investigate a set of features exhibited by the ensemble such as: a) the evolution of the KL matrix as a function of the number of training epochs, b) the evolution of the KL matrix as a function of the training dataset size, and c) the comparison between these behaviours and the standard loss function in Eq.~\eqref{eq:npe_loss}. 
Before presenting the results, we note that all the relevant technical details regarding the network architectures, training settings, prior ranges, etc., are provided in Appendix~\ref{sec:appendix_A}.

The main results for this subsection are presented in Fig.~\ref{fig:KL_GM_all} and Fig.~\ref{fig:corners}. 
They demonstrate a number of core properties that justify our statement that the KL matrix can be used as a diagnostic tool for ensembles of posterior approximations. 
A key use case is monitoring the KL matrix for a single ensemble during training. 
As shown in the left panel of Fig.~\ref{fig:KL_GM_all}, when the KL divergence stabilises, we can conclude that further training will not improve ensemble consistency, and the asymptotic KL value provides an estimate of the systematic uncertainty for that training configuration. 
To estimate the systematic uncertainty for a single, fixed training setup, only one ensemble of multiple members is required.

Firstly, if we look at the left panel of Fig.~\ref{fig:KL_GM_all}, we observe the behaviour of the average KL divergence in the KL matrix as a function of the training epoch for a number of different training set sizes. 
Specifically, we see that for all training set sizes, the KL divergences initially decrease as the networks learn a better approximation to the true posterior. 
Eventually, once the information contained in the training dataset has been fully utilised and the networks have converged, an asymptotic value of the KL divergences is reached. 
The key insight here is that this asymptotic value can be used to directly estimate the systematic error associated with the specific training configuration (i.e., training set size, training dynamics, etc.) via the interpretation discussed in Section~\ref{sec:kl_gaussian}. 
In our plots, for visual clarity, we show the average of the off-diagonal KL matrix elements. 
However, the primary diagnostic for assessing convergence or flagging outliers is the maximum pairwise KL divergence, $\max_{i \neq j} K_{ij}$, as this value provides the lower bound on the approximation error.

If we then look at the middle panel of Fig.~\ref{fig:KL_GM_all}, we see a more detailed picture of the asymptotic values reached in the KL matrices for each experiment. 
These demonstrate empirically a number of key ideas pointed out in Section~\ref{sec:theory}. 
This panel also demonstrates how the KL matrix can be used to diagnose the source of disagreement. 
We observe two distinct regimes: (i) a data-limited regime at small $n_\mathrm{Train}$, where the KL divergence decreases as $1/n_\mathrm{Train}$ (in line with Eq.~\eqref{eq:KL_expectation}), indicating that disagreement is driven by insufficient training data; and (ii) an optimisation-limited regime at large $n_\mathrm{Train}$, where the KL divergence plateaus. 
This floor reflects intrinsic variance from the training dynamics (e.g., random initialisation, stochastic gradient descent, architectural instabilities), showing that simply adding more data will not improve consistency. 
This saturation value can be used to compare different training strategies (e.g. learning rate scheduling, hyperparameter choices, etc.), with lower values indicating more stable performance, and to estimate the systematic error associated with the training dynamics itself. 
We emphasise that this information cannot be obtained when training only a single network, but must be bootstrapped via a trained ensemble.

The rightmost panel of Fig.~\ref{fig:KL_GM_all} illustrates the long-term values of the validation loss for the different training datasets. 
The main takeaway here is simply that whilst the loss function shows similar convergence as the dataset size increases, the behaviour is subtly different from the KL matrices. 
This justifies our statement that the diagnostic tests we suggest here provide \emph{complementary} information to the loss function itself.

Finally, for reference, in Fig.~\ref{fig:corners}, we show samples drawn from the three networks in the ensemble at the end of the training process, after conditioning to the same reference observation. 
The different panels correspond to different training dataset sizes, which are chosen to match Fig.~\ref{fig:KL_GM_all}. 
In all the panels, we show the true posterior in black. 
Once again, we can appreciate that, while all panels show qualitatively good agreement between the true posterior and the approximants, the KL matrix provides a quantitative measure to assess the accuracy and consistency of the approximation. We demonstrate the convergence of both the average, and maximum/minimum KL divergences across the ensemble in the Appendix.
While visual inspection of these 2D marginals is feasible, the KL matrix is particularly powerful in high-dimensional problems. 
The KL divergence captures the full multivariate disagreement in a single scalar per ensemble member pair, providing a quantitative and automated diagnostic that can be integrated into training pipelines and systematic uncertainty budgets.

As a final comment, we stress that while the current work only presents results for the two-dimensional Gaussian mixture model, we have followed the same procedure for other typical benchmark examples from the \texttt{sbibm} package~\cite{lueckmann2021benchmarking}, reaching similar conclusions.
    
\subsection{Testing model misspecification} 
\label{sec:NPE_KL_misspecification}

\begin{figure*}[ht]
    \centering
    \includegraphics[width=\linewidth]{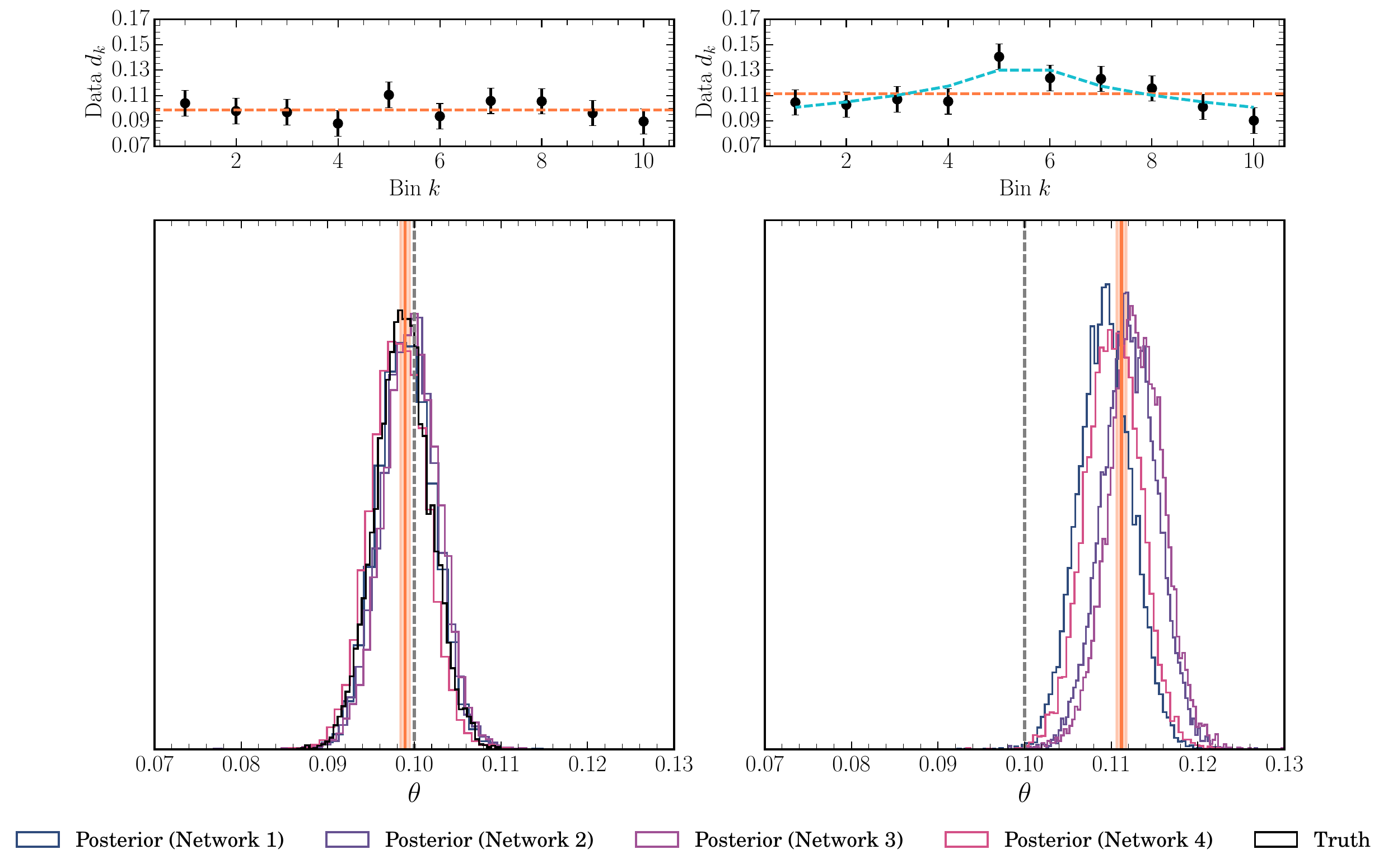}
    \caption{\textbf{Detecting Model Misspecification.} \emph{Left}: Posteriors (bottom plot) obtained from networks conditioned to an observation (top plot) compatible with the data used during the training. 
    In the top plot, the data realisation is shown in black, and the dashed orange line represents the input theoretical model $\mu_k$. 
    The orange vertical band in the bottom plot shows the maximal expected spread among estimates given the final value of the KL matrix ($\sim 7 \times 10^{-2}$) during training. 
    As expected, given the small KL at the end of the training, all posteriors agree. 
    For reference, we show the true posterior in black. 
    The dashed grey line represents the injected value of $\theta$ and the orange vertical line the mean value inferred from the data. 
    \emph{Right:} Posteriors (bottom plot) from the same networks conditioned to an observation (top plot) that is \emph{not} compatible with the data used during the training. 
    It is manifest that, once conditioned to an observation that does not match the model used in the training process, the spread in the posteriors increases significantly, which we can easily detect using the KL matrix. 
    The orange solid and grey dashed lines indicate the same quantities as in the left-hand figure.}
    \label{fig:non_matching}
\end{figure*}

\noindent In the previous section, we demonstrated the utility of the KL matrix as a diagnostic tool for tracking the evolution and stability of network training. 
In this section, we will further develop this idea and explore the possibility of using the KL matrix to detect model misspecification. 
"Misspecification" in our framework refers to the scenario where the simulator used for training does not accurately represent the data-generating process for the real observed data. 
This mismatch is revealed only when we apply the trained estimators to actual observations, as during training the model is correctly specified by definition. 
The underlying hypothesis is simple: given the flexibility of neural network models, we have no guarantees on the extrapolation behaviour of a trained model beyond data in the training set. 
In particular, for an ensemble of models, we cannot expect that if network predictions agree (as quantified by low values in the KL matrix, for example) within the training sample that they will each extrapolate in an identical manner. 
This is broadly a statement about the large degeneracy that exists in the parameter space of large neural networks when optimising a scalar loss function. 
In general, this is presented as a negative feature, however, here we point out that one can actually leverage this fact to detect model misspecification.

A crucial prerequisite for this test is to distinguish between training-induced variance and misspecification-induced variance. 
We define two quantities: (i) the training-time KL divergence, $\mathrm{KL}_\mathrm{train}$, which is measured on simulated test data from the same distribution as the training data, and quantifies the baseline disagreement due to the training process. 
(ii) The inference-time KL divergence, $\mathrm{KL}_\mathrm{obs}$, measured on the actual observed data. 
Our misspecification diagnostic is based on the difference $\Delta = \mathrm{KL}_\mathrm{obs} - \mathrm{KL}_\mathrm{train}$. 
A significant positive value of $\Delta$ suggests the observation is out-of-distribution, as it triggers enhanced disagreement among ensemble members compared to their baseline. 
Therefore, the test is most powerful when the ensemble has achieved a low, stable baseline $\mathrm{KL}_\mathrm{train}$, otherwise any signal from misspecification could be lost in the noise of training variance.

To test this hypothesis, we will train an ensemble of networks until the KL divergence among the networks in the ensemble is below a certain threshold value. 
Then, we will condition the density estimators on some observations that do not match the model in the simulator and compare the KL matrix on this new ensemble of posteriors to the values obtained during training. 
To systematically evaluate the performance of the KL divergence as a detector for model misspecification, we introduce a parameter to control the level of discrepancy between training and observed data. 
Following this procedure, we can keep track of the changes in the KL divergence between trained networks as we increase the level of model misspecification.

The specific example that we will study in this section is as follows: Each data realisation consists of $n = 10$ data points (in the following, we will label them with $k\in[1,n]$) drawn from a normal distribution with mean
\begin{equation}
    \mu_{k} = \theta + a \mathcal{B}_{k} \; , 
\end{equation}
and standard deviation $\sigma_{k} = \sigma = 0.01$. 
The parameter $\theta$ is drawn from a uniform prior $\theta \sim \text{Uniform}(-1, 1)$ and gives a constant contribution to the mean of the dataset and 
\begin{equation}
    \mathcal{B}_{k} \equiv 
\left( 1.01 - \left| \frac{2(k - 1)}{9} -1\right|^{0.25} \right) / \bar{\mathcal{B}_k} \; ,
\end{equation}
with $\bar{\mathcal{B}_k} \simeq 0.9326$, is a symmetric bump centred at $k = n/2$, with the denominator chosen to have $\max_{k\in [1,...,10]}(\mathcal{B}_{k}) = 1$. 
The contribution of this bump to $\mu_k$ is then controlled by the parameter $a$. 
Notice that for $a = 0$, there is no $k$ dependence in the dataset; i.e., the samples for different values of $k$ are independent realisations drawn from the same statistical distribution. 
For this very simple model, given $d_{k}$, we can determine the maximum likelihood estimator (and its variance) for the parameters analytically. 
In particular, in the absence of the bump, the solution is given by $\bar{\theta} = \sum_k d_{k} / n$ and $\sigma_{\bar{\theta}} = \sigma / \sqrt{n}$.

For the example presented in this section, we fix $N_d = 10000$ as the size of our training dataset\footnote{To be more accurate, as explained in~\cref{sec:appendix_A}, we draw $N_d$ values for $\theta$, which we keep fixed during the training process and, every time we show a pair $(\theta, d_{k})$ to the networks in the ensemble, we draw an independent realisation of $d_{k}$ given $\theta$. 
Following this procedure, we effectively show many instances of the data corresponding to the same parameters, which allows the networks to better learn the statistical properties of the data.} and fix $a=0$ for all examples used during the training process. 
The ensemble used for this example consists of $N_{\rm Net} = 5$ networks. 
After fixing the network architecture and training strategy, we train and monitor the evolution of the KL divergence among all networks in the ensemble. 
At the end of our run, the maximal value of the KL divergence is $\sim 7 \times 10^{-2}$, corresponding to a maximal mismatch of $\sim .37 \sigma$ in the peak position (see~\cref{sec:kl_metric} for the conversion from KL values to $\sigma$-levels). 
As already mentioned, for all datasets we used during the training process, we fix $a=0$, meaning we only train our networks on examples that do not contain the contribution due to the bump. 
In other words, if the observed data consisted solely of $\theta$, the candidate model would be correctly specified. 
This establishes a clear reference point against which the effects of misspecification can be assessed. 
Specifically, we can compare the observed values in the KL matrix for a given (potentially misspecified) observation and compare them to $7 \times 10^{-2}$, or look for posteriors that differ by more than $\sim .37 \sigma$.

\begin{figure*}
    \centering
    \includegraphics[width=0.9\linewidth]{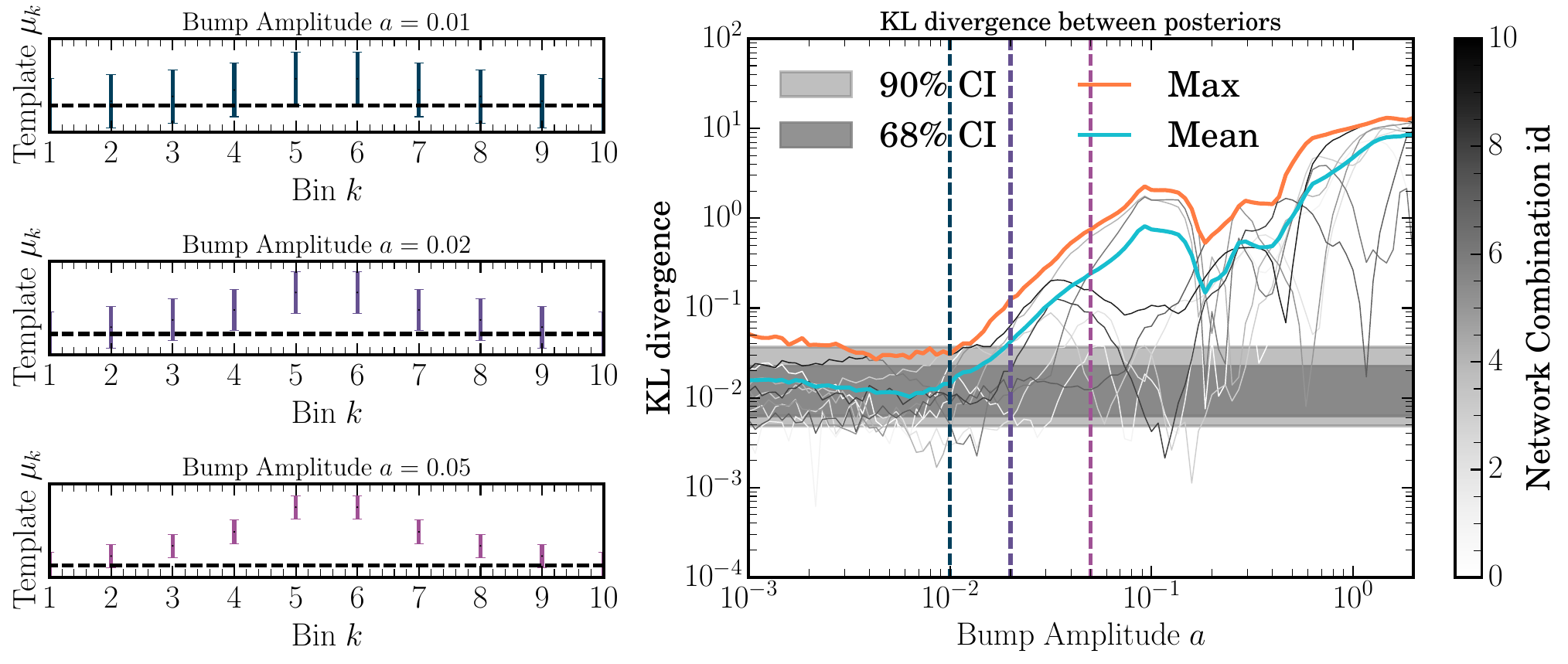}
    \caption{\textbf{Tracking the amount of model misspecification.} \emph{Left:} Three examples of the injection templates $\mu_k = \theta + a \mathcal{B}_k$, corresponding to the dashed vertical lines in the right-hand figure. 
    The error bars correspond to the $1\sigma$ noise levels in each bin, and the horizontal dashed lines indicate the equivalent flat template with $a = 0$. 
    \emph{Right:} The elements of the KL matrix as a function of the amplitude of the bump. 
    The grey bands are the percentiles computed among networks in the regime where the bump contribution is negligible compared to the flat component. 
    The remaining curves indicate the behaviour of the KL divergence associated with each network combination. 
    The mean (light blue) and maximum (orange) curves are specifically highlighted to be used as a diagnostic for detecting model misspecification.}
    \label{fig:kl_ampl_bump}
\end{figure*}

In Fig.~\ref{fig:non_matching}, we compare the results obtained conditioning the networks of the ensemble on an observation matching data used for the training process (left plot), i.e., $a = 0$, and an observation that is misspecified (right plot) with $a = 3 \times 10^{-2}$. 
The two plots in the first row show the data realisation with and without adding the bump component. 
It is evident that with the addition of the bump, the data deviates from the simulation model (which assumes a flat mean across the bins) used during the training process. 
In the bottom row, we show the posteriors obtained from the different networks in the ensemble after conditioning on the corresponding observation. 
Specifically, on the left hand side, which excludes the bump, all posteriors agree well (i.e., to the level expected from the KL matrix analysis) with each other and with the true posterior (shown in black). 
On the other hand, on the right hand side, we see a much larger spread among the posteriors given by the different networks. 
To be more quantitative, after conditioning on an observation that lies outside the training set, we find a KL divergence among the networks of the ensemble that significantly exceeds the threshold value used during the training process. 
We highlight this visually by showing the orange vertical contours, which denote the $\pm 0.37 \sigma$ variance expected from the in-sample training result. 
We can see that the posteriors in the misspecified model disagree to a level that is not captured by this error.

To better explore this behaviour, in the right plot of Fig.~\ref{fig:kl_ampl_bump}, we show how the KL divergences among the networks of the ensemble change as we condition on observations with increasingly large values of $a$ in the range $a \in [10^{-3}, 2]$. 
For reference, in the left plots, we show the relative size of the bump for a number of benchmark cases corresponding to the vertical lines in the right-hand figure. 
With these results, we can see clearly that the deviation in the values of the KL matrix starts to become visible when $a$ becomes larger than the threshold value $a_{\rm Thresh} \sim 1.3 \times 10^{-2}$. 
Interestingly, this corresponds to deviations in the data at about the 1-2$\sigma$ level. 
This suggests that our proposed misspecification test is highly sensitive. 
For values of $a$ well below $a_{\rm Thresh}$, we see that the values of the KL divergence remain consistent with the value achieved during the training process. 
However, for $a$ above $a_{\rm Thresh}$, the KL divergence starts to increase significantly. 
Thus, we can conclude that model misspecification can induce a significant increase in the KL divergence with respect to the value reached during training. 
In general, this provides a brand new test for detecting model misspecification that directly harnesses the power of ensemble learning.

Classical approaches to model misspecification detection include goodness-of-fit tests based on $\chi^2$ statistics comparing observed and expected data distributions, posterior predictive checks, and Bayesian model comparison via evidence ratios. 
In our specific example with the bump model, a $\chi^2$ test on the residuals between the data and the best-fit flat model would indeed detect the misspecification at comparable sensitivity levels ($\approx 1-2\sigma$). 
While classical tests like the $\chi^2$ statistic are computationally cheaper and should be a standard first check, our ensemble-based approach offers several complementary advantages: (i) it does not require explicit specification of a null hypothesis or alternative model, (ii) it provides a continuous diagnostic that can be monitored during analysis rather than requiring a separate validation step, and (iii) it naturally accounts for parameter uncertainty through the ensemble disagreement rather than relying on point estimates. 
Recent ML-enhanced approaches include using classifier-based test statistics~\cite{cranmer2016approximating}, where a neural network is trained to distinguish between data generated under different hypotheses. 
Our KL-based approach is conceptually related but operates in posterior space rather than data space, making it naturally integrated with the inference procedure itself. 
A systematic comparison of these approaches across various misspecification scenarios would be valuable future work.

\section{Outlook and conclusions}
\label{sec:conclusions}

\noindent In this work, we have explored the usage of ensemble learning as a tool to diagnose convergence, stability, and model fidelity in Simulation-Based Inference (SBI). 
The core idea is very simple: if an SBI algorithm reliably approximates the true posterior distribution, then independently trained models on the same simulation budget should yield consistent results. 
In other words, since the true posterior is unique, a small discrepancy among ensemble members is a necessary, but not sufficient, condition to ensure that all posterior estimates can be sufficiently close to the true posterior. 
Therefore, disagreement among ensemble members, as quantified by the KL divergence in this work, becomes an accessible and informative signal of failure modes such as undertraining, optimiser variance, or model misspecification.

We have shown that the KL divergence between posterior approximations within an ensemble provides a principled, quantitative diagnostic without requiring access to the true posterior. 
The KL matrix constructed from pairwise divergences among ensemble members encodes information about convergence properties, with elevated divergences signalling internal inconsistency and potential unreliability of the inferred posterior for a specific observation. 
In particular, we demonstrated that as the size of the training dataset increases, the KL divergences among the ensemble members decrease until they reach a floor set by the intrinsic variance of the training procedure and the network architecture, which offers a useful lower bound on the systematic errors intrinsic to the training process.

Moreover, we demonstrated in Section~\ref{sec:KL_applied_to_ensemble} that ensemble disagreement, as measured by the KL divergence, can serve as a practical detector for model misspecification. 
Conditioning trained estimators on out-of-distribution observations results in a measurable increase in divergence between ensemble members, underscoring the diagnostic power of this approach beyond mere convergence assessment. 
Our experiments with controlled perturbations to observed data confirmed that the KL divergence can track systematic deviations beyond the information provided by the loss function. 
It would be interesting in future work to perform a systematic comparison of the sensitivity of this KL-based test against classical techniques and other ML-enhanced tools. 
Another interesting direction would be to deliberately train on misspecified simulations (e.g., with known simplifications) and use the KL diagnostic to assess when and where these simplifications become problematic. 
It would also be interesting to explore further ways to enhance ensemble learning for SBI. 
For example, by drawing inspiration from parallel tempering, training models with different learning rates, allowing occasional information exchange via the KL matrix could improve robustness and generalisation, as well as achieving a better balance between exploration and exploitation. 
Furthermore, in analogy with Reversible Jump MCMC (RJ-MCMC), dynamically adapting model architectures within an ensemble would allow models to “jump” between different complexities based on performance, potentially leading to more efficient use of computational resources. 
Beyond leveraging ensembles for consistency checks, combining outputs from diverse models also offers a natural path to more stable posterior estimates by mitigating variance across architectures.

Overall, our results suggest that ensemble training, combined with internal divergence metrics such as the KL matrix, provides a scalable, flexible, and inference-agnostic framework for improving the transparency and reliability of SBI pipelines. 
This paradigm is particularly well suited for scientific settings where inference is performed once, but scrutiny must be rigorous. 
Future work will explore generalisations to other divergence measures, combinations with simulation-based calibration methods, and applications to high-dimensional real-world inference tasks.

\section*{Data Availability}
\noindent The code to reproduce all results in this manuscript is publicly available on GitHub at \url{https://github.com/james-alvey-42/sbi-ensemble-diagnostics}, and the datasets generated and analysed during the current study are available in the Zenodo repository, \url{https://zenodo.org/records/17522623}.

\section*{Acknowledgements}
\noindent We thank Jonas Elias El Gammal, Michele Mancarella, Michael Williams, and Christoph Weniger for their comments on a nearly final version of this draft. 
MP would like to thank Lorenzo Cevolani for very useful comments on a draft of this work. 
MP and JA acknowledge the hospitality of Imperial College London, which provided office space during some parts of this project. 
JA is supported by a fellowship from the Kavli Foundation. 
The work of MP is supported by the Comunidad de Madrid under the Programa de Atracción de Talento Investigador with number 2024-T1TEC-3134.

\bibliography{main}

\appendix

\renewcommand{\thesection}{\Alph{section}}
\crefname{section}{Appendix}{Appendices}

\section{Further details and tests}
\label{sec:appendix_A}

\noindent In this appendix, we provide further details on the setups used for this work, along with additional plots to clarify the training process and the behaviour of the ensembles.

All posterior density estimation performed in this work relies on the Neural Spline Flow (NSF) architectures~\cite{Durkan:2019aaa} implemented via the \texttt{build\_nsf} function\footnote{For all examples considered in this work, we stick to the default internal architecture parameters for components such as the number of coupling layers, hidden features, and spline configurations.} in the \texttt{sbi} library~\cite{boelts2024sbi}. 
Each NSF network models the conditional posterior distribution $p(\theta|x)$, where simulated observations $x$ are provided as input and the network outputs a flexible approximation over parameters $\theta$. 
Training data $x$ is generated on the fly and loaded in mini-batches of size 64. 
A particular aspect of our training process that may be non-standard is ``noise resampling". 
While the set of parameter values $\{\theta\}$ is kept fixed throughout training, the observations $\{x\}$ are dynamically regenerated by the simulator whenever data is requested. 
This approach effectively augments the training data with new noise realisations for each $\theta$, enhancing robustness and reducing variance induced by fixed realisations.

For all our examples, we independently train an ensemble of networks, each with an input dimension matching the observation space of the specific simulation task. 
Each network is optimised using the \texttt{AdamW} optimiser with an initial learning rate of $10^{-2}$. 
A \texttt{StepLR} scheduler reduces the learning rate by a factor of $0.9$ every $10$ epochs. 
As detailed in the main text, a central component of our ensemble approach is the use of the KL divergence to assess agreement among the ensemble members. 
Specifically, every $10$ epochs, we estimate the pairwise KL divergence between the learned posterior distributions using $10,000$ samples drawn from the estimated posteriors. 
Early stopping is triggered when the maximum absolute KL divergence across all pairs falls below a tolerance of $10^{-3}$, indicating that the networks have converged to mutually consistent posterior approximations. 

The Gaussian Mixture Model example used in Section~\ref{sec:NPE_KL_convergence} follows the default configuration from the \texttt{sbibm} benchmark library~\cite{lueckmann2021benchmarking} to ensure reproducibility. 
The specific model details are as follows:
\begin{itemize}
    \item The prior distributions for both parameters are uniform: $\theta_1, \theta_2 \sim \text{Uniform}(-10, 10)$.
    \item The likelihood $p(x|\theta)$ is a mixture of two 2D Gaussian distributions with means at $(\theta_1, \theta_2)$ and $(-\theta_1, -\theta_2)$, with equal weights of 0.5.
    \item The standard deviations of each mode are as specified in the \texttt{sbibm} benchmark library implementation~\cite{lueckmann2021benchmarking}.
\end{itemize}

\begin{figure*}
    \centering
    \includegraphics[width=0.85\linewidth]{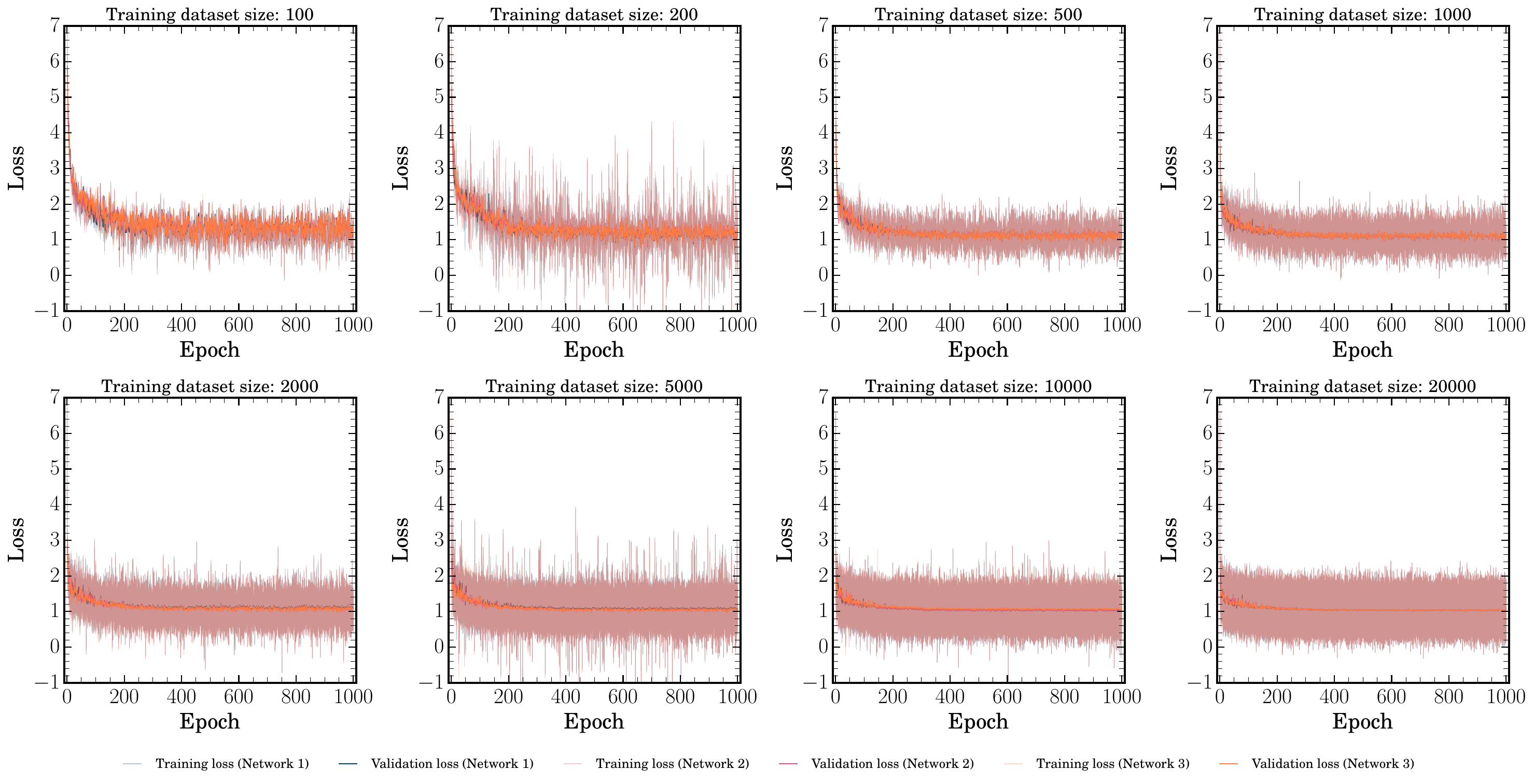}
    \caption{\textbf{Training and Validation Losses.} Training (light colour) versus validation loss (full colour) for the different dataset sizes (100, 200, 500, 1000 top row, left to right, and 2000, 5000, 10000, 20000 left to right bottom row) as a function of epochs for all the networks in the ensemble (red, green, blue) discussed in Section~\ref{sec:NPE_KL_convergence}.}
    \label{fig:losses}
\end{figure*}

\begin{figure*}
    \centering
    \includegraphics[width=0.85\linewidth]{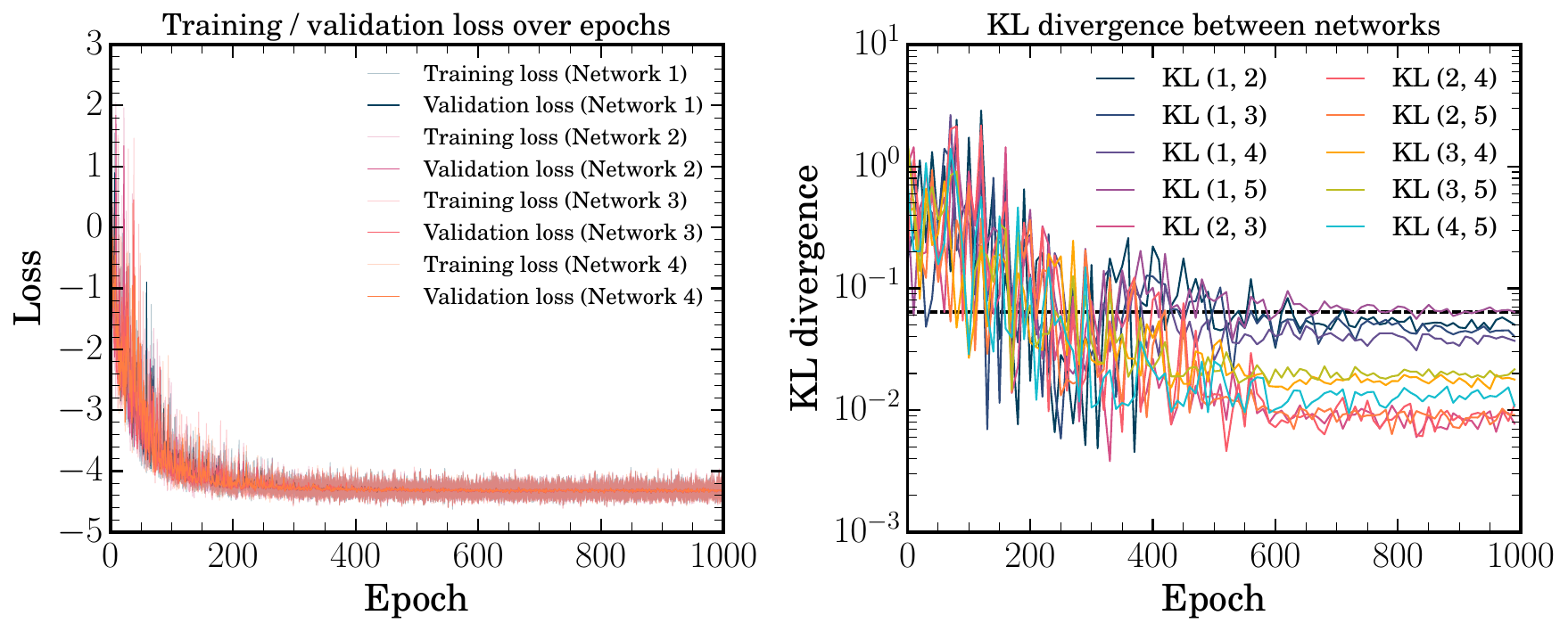}
    \caption{\textbf{Training Losses and KL matrix.} \emph{Left:} Training and validation loss as functions of training epochs for the five networks in the ensemble trained in Section~\ref{sec:NPE_KL_misspecification} (see legend). 
    \emph{Right:} Evolution of the KL divergence between network combinations in the ensemble (see legend for various combinations) as a function of training epochs.}
    \label{fig:KL_losses_extrapolation}
\end{figure*}

\begin{figure*}
    \centering
    \includegraphics[width=0.5\linewidth]{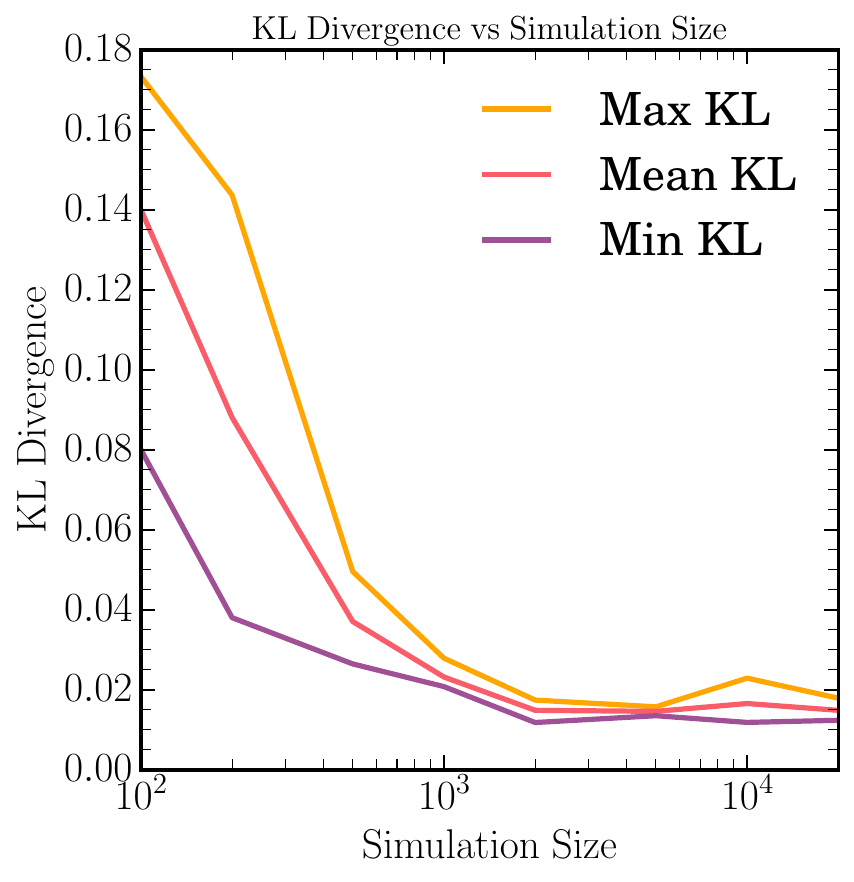}
    \caption{\textbf{Evolution of maximum/average KL divergences.} Behaviour of the maximum, minimum and average values of the KL divergences in the ensembles trained for the example discussed in Section~\ref{sec:NPE_KL_convergence}. We see that all three quantities decrease systematically as the dataset size increases until convergence is reached.}
    \label{fig:max_KL}
\end{figure*}

We also provide further details as to the ensembles used in Section~\ref{sec:NPE_KL_convergence}. 
As explained in the main text, the ensembles consist of three networks that are trained in parallel using different training dataset sizes $n_{\rm{Train}}$, ranging from $100$ to $20000$ across the various experiments. 
We train each network for up to $1000$ epochs, fixing the validation set size to $n_{\rm{Val}}=2 \times n_{\rm{Train}}$. 
In Fig.~\ref{fig:losses}, we show the training/validation losses as functions of epochs. 
All panels, meaning all ensembles with different values of $n_{\rm{Train}}$, exhibit a similar, and expected, behaviour. 
After a steep decrease in the initial epochs, the training and validation losses oscillate randomly around some asymptotic value. 
While for small $n_{\rm{Train}}$, both training and validation loss exhibit quite a large variance around such asymptotic value, for larger $n_{\rm{Train}}$, the validation loss variance is strongly reduced. 
This behaviour originates from our choice for $n_{\rm{Val}}$: for larger $n_{\rm{Train}}$, the validation dataset is larger, and the results are less subject to statistical fluctuations.

We conclude this appendix by providing further details on the models used in Section~\ref{sec:NPE_KL_misspecification}. 
As already mentioned in the main text, for this example, parameters $\theta$ are uniformly sampled from $[-1, 1]$. 
The corresponding observations $x$ are produced by a simulator that adds Gaussian noise ($\sigma=0.01$) to $\theta$ across the 10 independent measurements. 
In practice, each observation corresponds to a 10-dimensional data vector modelling noisy measurements. 
The ensemble used for this example consists of five networks trained in parallel using the procedure described above in this appendix. 
The training dataset consists of 10000 $\theta$ samples (the observations are resampled dynamically as discussed above), and the validation dataset has size 2000. 
In the left panel of Fig.~\ref{fig:KL_losses_extrapolation}, we show the training/validation losses for the five networks in the ensemble as a function of training epochs. 
As expected, after a steep initial decrease, the losses of the five networks flatten out and fluctuate randomly around an asymptotic value. 
In the right panel of the same figure, we show the evolution of the KL divergences among the different networks in the ensemble as functions of training epochs. 
Again, after an initial phase, they all tend to scatter around some nearly constant value. 
We notice that all KL divergences involving network ID 1 are significantly larger, indicating that there is less consistency between this particular model and the others in the ensemble. 
For reference, the maximum value of KL among all combinations at the end of the training process is $\simeq 7 \times 10^{-2}$. 
We use this number in the main text as the lower bound to which we compare detections of model misspecification.

\end{document}